# Evidence chain for time-reversal symmetry-breaking kagome superconductivity

**Authors:** Hanbin Deng[1]*, Guowei Liu[1]*, Z. Guguchia[2]*, Tianyu Yang[1]*, Jinjin Liu[3,4]*, Zhiwei Wang[3,4]†, Yaofeng Xie[5], Sen Shao[6], Haiyang Ma[7], William Liège[8], Frédéric Bourdarot[9], Xiao-Yu Yan[1], Hailang Qin[7], C. Mielke III[2], R. Khasanov[2], H. Luetkens[2], Xianxin Wu[10], Guoqing Chang[6], Jianpeng Liu[11], Morten Holm Christensen[12], Andreas Kreisel[12], Brian Møller Andersen[12], Wen Huang[13], Yue Zhao[1], Philippe Bourges[8], Yugui Yao[3,4], Pengcheng Dai[5], Jia-Xin Yin[1,7]†

**Affiliations:**

[1]Department of Physics, Southern University of Science and Technology, Shenzhen, Guangdong, China.

[2]Laboratory for Muon Spin Spectroscopy, Paul Scherrer Institute, CH-5232, Villigen PSI, Switzerland.

[3]Centre for Quantum Physics, Key Laboratory of Advanced Optoelectronic Quantum Architecture and Measurement (MOE), School of Physics, Beijing Institute of Technology, Beijing, China.

[4]Beijing Key Lab of Nanophotonics and Ultrafine Optoelectronic Systems, Beijing Institute of Technology, Beijing, China.

[5]Department of Physics and Astronomy, Rice University, Houston, Texas 77005, USA.

[6]Division of Physics and Applied Physics, School of Physical and Mathematical Sciences, Nanyang Technological University, Singapore 637371, Singapore.

[7]Quantum Science Center of Guangdong-Hong Kong-Macao Greater Bay Area (Guangdong), Shenzhen, China.

[8]Université Paris-Saclay, CNRS-CEA, Laboratoire Léon Brillouin, 91191, Gif sur Yvette, France.

[9]Université Grenoble Alpes, CEA, INAC, MEM MDN, F-38000 Grenoble, France.

[10]CAS Key Laboratory of Theoretical Physics, Institute of Theoretical Physics, Chinese Academy of Sciences, Beijing, China.

[11]School of Physical Science and Technology, ShanghaiTech University, Shanghai 201210, China.

[12]Niels Bohr Institute, University of Copenhagen, DK-2200 Copenhagen, Denmark.

[13]Shenzhen Institute for Quantum Science and Engineering, Southern University of Science and Technology, Shenzhen 518055, Guangdong, China.

*These authors contributed equally to this work.

†Corresponding authors. E-mail: zhiweiwang@bit.edu.cn; yinjx@sustech.edu.cn

**Superconductivity and magnetism are antagonistic quantum matter, while their intertwining has long been considered in frustrated-lattice systems[1-3]. In this work, we utilize scanning tunneling microscopy and muon spin resonance to discover time-reversal symmetry-breaking superconductivity in kagome metal $Cs(V,Ta)_3Sb_5$, where the Cooper pairing exhibits magnetism and is modulated by it. In the magnetic channel, we observe spontaneous internal magnetism in a full-gap superconducting state. Under perturbations of inverse magnetic fields, we detect a time-reversal asymmetrical interference of Bogoliubov quasi-particles at a circular vector. At this vector, the pairing gap spontaneously modulates, which is distinct from pair density waves occurring at a point vector and consistent with the theoretical proposal of unusual interference effect under time-reversal symmetry-breaking. The correlation between internal magnetism, Bogoliubov quasi-particles, and pairing modulation provides a chain of experimental clues for time-reversal symmetry-breaking kagome superconductivity.**

A kagome lattice is a lattice made of corner-sharing triangles. Initial research on kagome physics starts with its unusual quantum magnetism: the geometrical spin frustration can lead to the absence of a magnetic transition[4, 5], and geometrically localized electrons can lead to flat-band ferromagnetism[6]. Early considerations[1-3] of superconductivity in materials hosting kagome lattices, intriguingly, also show an intimate relationship with magnetism, including the concepts of ferromagnetic superconductors and time-reversal symmetry-breaking (TRSB) superconductivity. Recent studies of topological kagome magnets and superconductors further push the interplay between magnetism and correlations in the kagome lattice to the frontier of quantum materials[7]. Particularly, research on the $CsV_3Sb_5$ class of kagome superconductors has widely discussed a TRSB charge order[8-15]. However, the nature of their superconductivity ground state remains elusive. In this work, we report the discovery of TRSB superconductivity in the kagome metal $Cs(V,Ta)_3Sb_5$ via both magnetic and electronic probes.

**Full pairing gap and spontaneous internal magnetism**

The kagome superconductor[16] $CsV_3Sb_5$ crystallizes into the $P6$/mmm space group, with a kagome network of vanadium cations coordinated by octahedra of Sb that is further separated by layers of Cs (inset of Fig. 1**a**). When 14% Ta atoms are doped into the V-kagome layer[17,18], the charge density wave order is fully suppressed and the critical temperature of superconductivity $T_C$ is enhanced to 5K. We focus our study on the Sb surface, which tightly bonds to the kagome layer and is one of the natural cleavage surfaces. Imaging at high bias voltages reveals the underlying Ta dopants, the counting of which is consistent with the nominal doping concentration (Fig. 1**a**). These impurities are nonmagnetic and are the major scattering source for quasi-particles. Imaging at low bias voltages shows the individual Sb atoms (Fig. 1**b**), and the corresponding Fourier transform confirms the absence of 2×2 charge order (Fig. 1**c**). Probing the differential conductance deep in the superconducting state, we observe a fully opened energy gap (Fig. 1**d**). This gap disappears at $T_C$, and its lineshape at 30mK fits with a BCS gap function (Fig. 1**e**), both of which demonstrate it as a Cooper pairing gap. The sharp coherence peaks located at ±0.86 meV define their pairing gap size $\Delta_{\pm}$.

The full pairing gap is further confirmed by the muon spin resonance (μSR), which is a magnetic-sensitive probe[19]. We perform transverse-field muon spin resonance experiments down to 20mK with a field of 10mT applied along the $c$-axis (See SI for more details). By extracting the first and second moments of the inhomogeneous field distribution from the muon spin depolarization rate, we obtain the temperature evolutions of diamagnetism signal $B_{\rm dia}(T)$ and the inverse square of the in-plane magnetic penetration depth $\lambda_{ab}^{-2}(T)$ (a fundamental property that is proportional to the superfluid density), respectively (Fig. 1**f**). Both signals emerge below $T_C$, indicating the bulk character of superconductivity. Notably, $\lambda_{ab}^{-2}$ reaches its zero-temperature value exponentially, demonstrating full gap superconductivity.

After confirming the full gap, we employ the zero-field μSR experiments to probe whether there is TRSB in the superconducting state. The inset of Fig. 1**g** displays the zero-field μSR spectrum, measured at $T$ = 1.5 K. Since the relaxation is decoupled by a small external magnetic field (50 G) applied longitudinally to the muon spin polarization, the zero-field relaxation is therefore due to spontaneous fields which are static on the microsecond timescale[20]. Figure 1**g** plots the internal field width $\Gamma(T)$, which shows a noteworthy increase upon lowering the temperature below $T_C$. This observation indicates the enhanced spread of internal fields sensed by the muon ensemble concurrent with the onset of superconductivity. The increase in the relaxation below $T_C$ is estimated to be 0.012 $\mu s^{-1}$, corresponding to a characteristic field strength $\Gamma/\gamma_\mu$=0.15G, where $\gamma_\mu$ is the gyromagnetic ratio of the muon. This is comparable to what has been observed in superconductors that are believed to be TRSB, such as $Sr_2RuO_4$[20]. A similar enhancement of $\Gamma$ below $T_C$ has been observed[14,15] in $AV_3Sb_5$ (A=K, Rb, Cs) when the charge order is suppressed by pressure (Fig. 1**h**). However, not limited to kagome superconductors, the electronic feature of TRSB superconductivity has long been elusive.

**Time-reversal asymmetrical interference of Bogoliubov quasi-particles**

To resolve the tiny TRSB signal from the electronic structure, we designed magneto-electronic interference experiments at 0.3K running for four months. The key fermiology of this kagome superconductor, which consists of an inner $\alpha$ band and outer $\beta$ band(s), is shown in Fig. 2**a**. Three dominant backscattering vectors ($\boldsymbol{Q}_\alpha$, $\boldsymbol{Q}_\beta$, and $\boldsymbol{Q}_{\alpha\beta}$) arising from these two sets of Fermi surfaces are seen by our experiments, as shown below. We collect the tunneling conductance g($\boldsymbol{r}$, $E$) under different magnetic fields for a large field of view (~ 500Å × 500Å). We then map a ratio[21,22] Z($\boldsymbol{r}$, $E$)=g($\boldsymbol{r}$, +$E$)/g($\boldsymbol{r}$, –$E$), the process of which selects the Bogoliubov quasi-particle interference featuring a particle-hole symmetry. We collect these maps at $E$ = 0.75Δ with increasing c-axis magnetic field up to above its critical field (1.5T). Through the Fourier transformation of these maps, we obtain the field-dependent Bogoliubov quasi-particle interference Z($\boldsymbol{q}$) in Fig. 2**b**. They show a progressive increment of Bogoliubov quasi-particle scatterings at three $\boldsymbol{Q}$ vectors with increasing field.

Based on the magnetic sensitivity of Z($\boldsymbol{q}$), we further design an interference experiment with an innovative TRSB-sensitive set-up as shown in Fig. 2**c**. We collect $Z(\boldsymbol{q})$ data with opposite $c$-axis fields +$B$ and -$B$. Then, we look at their signal difference (δZ($q$)=Z($q$,+$B$)-Z($q$,-$B$)), which defines the time-reversal asymmetrical interference of Bogoliubov quasi-particles. In the middle ($E$ = 0.75Δ) and right ($E$ = 0.25Δ) panels of Fig. 2**d**, we indeed observe a TRSB signal δZ($\boldsymbol{q}$) at a circular vector $\boldsymbol{q} = \boldsymbol{Q}_\alpha$, under $B$ = ±0.5T. We note that δZ/Z = 12±2%, which is orders of magnitude larger than the random noise of Z signal on the level of 0.1% to 1%. This TRSB signal disappears at energies outside the pairing gap (Fig. 2**d** left, $E$ = 1.25Δ), at the critical magnetic field (Fig. 2**e** left), and at $T_C$ (Fig. 2**e** right) (See SI for more data and discussions). We further explore its behavior with reversed in-plane fields $B$ = ±1T applied along two high-symmetry directions: Γ-M and Γ-K (Figs. 2**f-h**). In the obtained δZ($\boldsymbol{q}$) data taken at the energy outside the superconducting gap, we reconfirm the absence of δZ($\boldsymbol{q}$) signal. Inside the superconducting gap, we observe that the time-reversal asymmetric signal at $\boldsymbol{Q}_\alpha$ again emerges. In addition, different from a nearly isotropic signal as in the $c$-axis field case, the δZ(q) signal becomes two segmented arcs, which is owing to the additional Doppler shift effect of in-plane field-induced screening supercurrent[23]. The TRSB setup, as well as the energy, magnetic field, and temperature dependences of the δZ($\boldsymbol{q}$) signal, attest to its intimate relation with the internal magnetism of superconductivity. Our experiment is also in line with the nonreciprocal transport and superconducting diode phenomena observed in kagome-superconductor-based devices[24-26]. These TRSB-related experiments suggest rich interplay between internal magnetism and Cooper pairs, and their response to the magnetic or electrical field applied along reverse directions (with a component parallel and antiparallel to the internal magnetism direction) can be different.

**Spontaneous Cooper pairing modulation**

Although both experiments support the TRSB of the superconducting state, we also note a crucial difference between the internal field observation and time-reversal asymmetrical Bogoliubov quasi-particle interference. The former is detected under zero field while the latter is detected under reversed magnetic fields, and we try to fill this gap by designing a new challenging experiment for their possible connection. As the internal magnetism is spontaneous, the magnetism-sensitive Cooper pairing is also expected to exhibit spontaneous modifications at the aforementioned TRSB scattering channel[27-29]. For instance, a recent theory[28] has suggested that under TRSB, the nonmagnetic impurity can cause pairing gap modulations distinct to the pair density wave occurring at a point vector. The decisive way to check the pair modulation is by measuring the gap map. We find a large clean Sb surface (400Å × 400Å, Fig. 3**a**) and map the energy of superconducting coherence peaks at both negative bias voltage $\Delta_{-}(\boldsymbol{r})$ and positive bias voltage $\Delta_{+}(\boldsymbol{r})$ for the same field of view at 0.3K (Figs. 3**b**-**d**). It is clear that the $\Delta_{+}(\boldsymbol{r})$ map strongly mimics the $\Delta_{-}(\boldsymbol{r})$ map, confirming the particle-hole symmetry of Cooper pairing at the atomic scale. Then we obtain the gap map $\Delta(\boldsymbol{r}) = [\Delta_{+}(\boldsymbol{r}) - \Delta_{-}(\boldsymbol{r})]/2$ in Fig. 3**e**, showing detectable modulations. The Fourier transform of the gap map as $\Delta(\boldsymbol{q})$ is shown in Fig. 3**f**, which exhibits pronounced signals at the circular vector $\boldsymbol{q} = \boldsymbol{Q}_{\alpha}$. At this circular vector, its intensity exhibits a small anisotropy with stronger intensity along Bragg peak directions, similar to that of the $\delta Z(\boldsymbol{q})$ signal. We further elucidate the real-space features of the Cooper pairing modulation by performing an inverse Fourier transform for $\Delta(\boldsymbol{q})$ at $\boldsymbol{Q}_{\alpha}$ in Fig. 3**g**. The real-space modulations contributing to the signal at $\boldsymbol{Q}_{\alpha}$ in $\Delta(\boldsymbol{q})$ are rather random, likely arising from the interplay between underlying randomly distributed nonmagnetic dopants and the TRSB pairing state. While quasi-particle interference is often detected in several cuprates and iron-based superconductors, the gap modulations at similar vectors have been missing. Therefore, both $\delta Z(\boldsymbol{q})$ and $\Delta(\boldsymbol{q})$ signals can serve as electronic fingerprints of TRSB superconductivity.

**Discussion and conclusion**

It is striking to observe $\delta Z(\boldsymbol{q})$ and $\Delta(\boldsymbol{q})$ emerging predominately at the same vector $\boldsymbol{Q}_{\alpha}$ even though they are from independent measurements, which supports their common relation with the underlying internal magnetism. The former regards time-reversal asymmetrical modulation of the Bogoliubov states with perturbations from inversed magnetic fields, while the latter regards the spontaneous modulation of the Cooper pairing strength. Thus, spontaneous internal magnetism, Bogoliubov quasi-particle interference, and Cooper pairing modulation are intertwined in our experiments. They establish an evidence chain for the TRSB superconductivity in this kagome metal at the experimental level. This finding is complementary to previous detection of TRBS charge order, advancing our knowledge of emergent orders featuring magnetic-electronic duality in the kagome lattice. In reference to our first-principles calculations, $\boldsymbol{Q}_{\alpha}$ is related to the intra-band scattering of α band, which is mainly composed of V/Ta $d_{xz/yz}$ orbitals and Sb $p_z$ orbitals (from the Sb atom within the kagome layer). Its association with TRSB is beyond consideration in a simplified kagome lattice[31-36] but is in line with the important role of *p*-*d* orbital hybridization in the electronic correlation discussed in kagome superconductors[37-40] as well as in copper-based and iron-based superconductors[41-44]. Microscopically, our results also suggest that the full gap superconductivity in this kagome metal may host a two-component TRSB superconducting order parameter $\Delta_1$+i$\Delta_2$ (such as *s*+i*s*, *p*+i*p*, and *d*+i*d*). Under such an order parameter, impurities can induce surrounding supercurrents that form internal magnetization and modulate the pairing gap. An external magnetic field can couple with the vectorial magnetization and affect the distribution of supercurrents, modifying the superconducting electronic structures. This may further result in time-reversal asymmetrical interference of Bogoliubov quasi-particles. Given the research trend that there lack of a well-accepted solid-state example for the simplest TRSB *p*-wave superconductivity and order parameters for canonical TRSB superconductor candidates including $UTe_2$ and $Sr_2RuO_4$ remain elusive, we expect substantial theoretical efforts are required in building a model for TRSB kagome superconductivity as constrained by our experiments. Crucially, our experimental work utilizing cutting-edge techniques builds up the correspondence between internal magnetism, Bogoliubov quasi-particles, and Cooper pairing, providing a powerful methodology for revealing TRSB superconductors.

## Figures

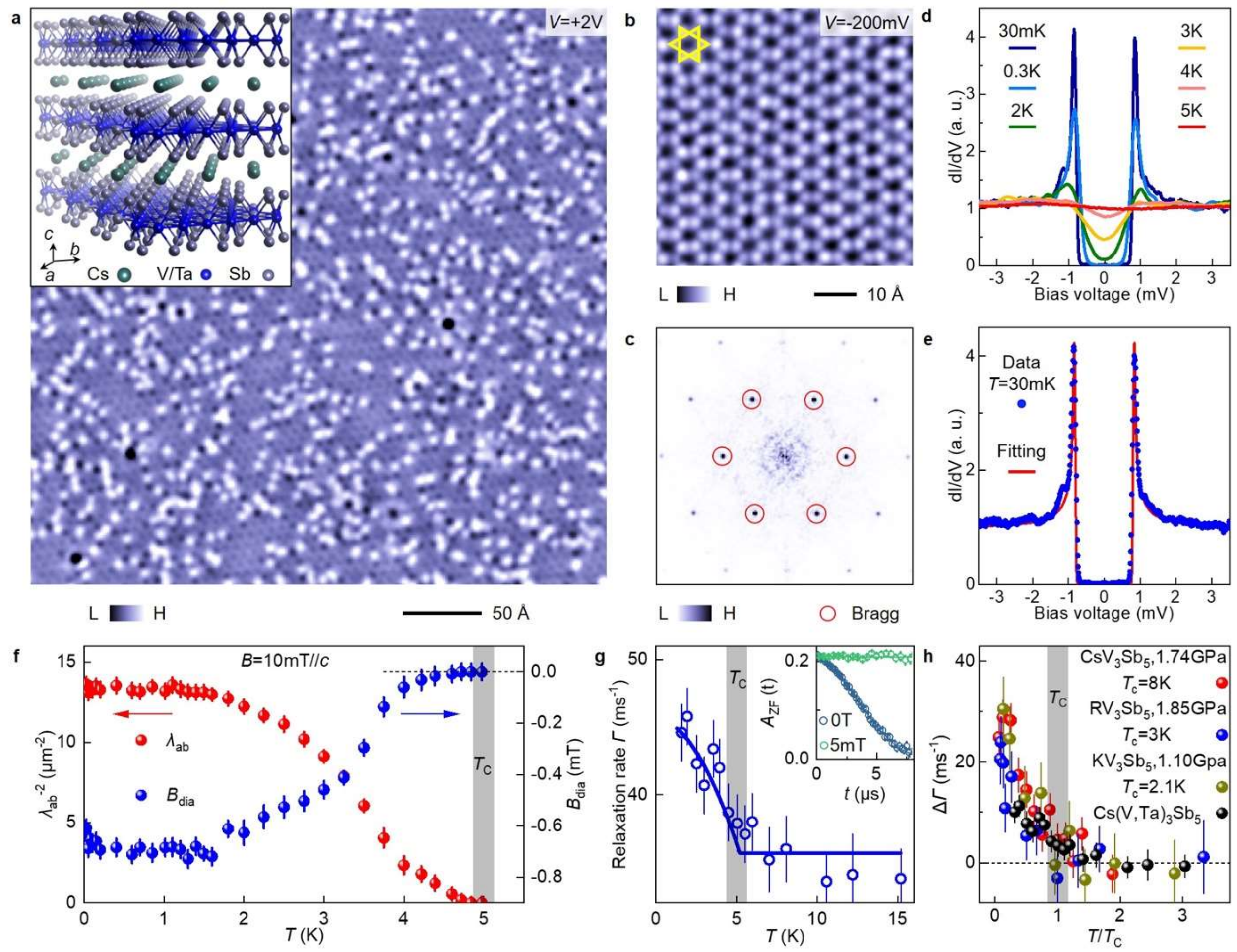

**Figure 1 Full pairing gap and spontaneous internal magnetism. a,** A topographic image of Sb surfaces taken at large bias voltage ($V$ = 2V, $I$ = 200pA, $T$ = 0.3K), showing underlying Ta dopants at bright spots. The inset shows the crystal structure of the kagome superconductor $Cs(V,Ta)_3Sb_5$. **b,** Topographic image showing individual Sb atoms ($V$ = -200mV, $I$ = 200pA, $T$ =0.3K). The yellow lines illustrate the underlying kagome lattice. **c,** Fourier transform of the topographic data in (**b**), demonstrating the absence of 2 × 2 charge density wave order. Red circles mark the Bragg peaks. **d,** Differential conductance spectrum taken at different temperatures ($V$ = 5mV, $I$ = 1nA). **e,** Fitting the 30mK tunneling data with the BCS gap function. **f,** The temperature dependence of diamagnetic shift $B_{dia}$ and inverse square of the in-plane penetration depth $\lambda_{ab}^{-2}$ from the transverse field μSR. The magnetic field of 10mT is applied along $c$-axis. **g,** Temperature dependence of the zero-field muon spin relaxation rate, showing the spontaneous appearance of internal magnetism. The solid lines represent a heuristic guideline. The inset shows the μSR time spectra measured in zero-field (dark blue color) and under a small external magnetic field of 50G applied in a direction longitudinal to the muon spin polarization (green color). **h,** The absolute change of the electronic relaxation rate $\Delta\Gamma = \Gamma(T)$ - $\Gamma(T > T_C)$ for Ta doped $CsV_3Sb_5$ at ambient pressure, $KV_3Sb_5$ at $p$ = 1.1 GPa, $RbV_3Sb_5$ at $p$ = 1.85 GPa, and $CsV_3Sb_5$ at $p$ = 1.74 GPa, plotted as a function of normalized temperature $T/T_C$. The data for pressured kagome superconductors are extracted from Ref. 14 and Ref.15.

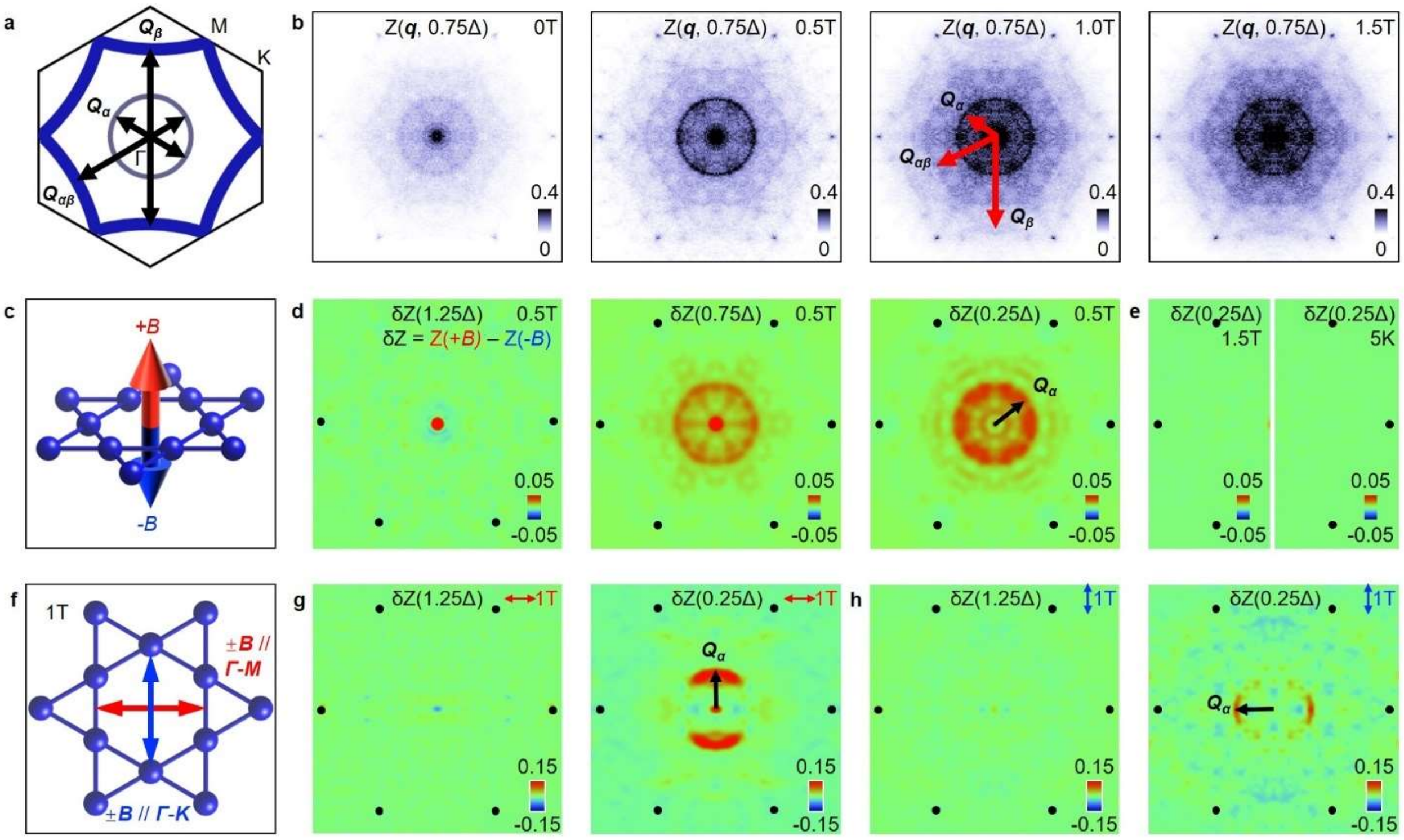

**Figure 2 Time-reversal asymmetrical Bogoliubov quasi-particle interference. a,** Fermiology of this kagome superconductor, which consists of the inner *α* band and the outer *β* band(s). We illustrate three main scattering vectors: $\boldsymbol{Q}_{\alpha}$, $\boldsymbol{Q}_{\beta}$ and $\boldsymbol{Q}_{\alpha\beta}$. (**b**) Bogoliubov quasi-particle interference Z($\boldsymbol{q}$) at different magnetic fields applied along the *c*-axis. The intensities at $\boldsymbol{Q}_{\alpha}$, $\boldsymbol{Q}_{\beta}$ and $\boldsymbol{Q}_{\alpha\beta}$ are all progressively increasing with increasing field strength. Data are six-fold symmetrized. **c,** Illustration of inverse magnetic fields applied perpendicular to the kagome lattice as a time-reversal perturbation of superconductivity. **d,** Time-reversal asymmetrical Bogoliubov quasi-particles interference signal δZ for $E$ = 1.25Δ (left), $E$ = 0.75Δ (middle), and $E$ = 0.25Δ (right), obtained by the subtraction of Z taken with opposite fields. The signal disappears outside the superconducting gap. Data are six-fold symmetrized. The black dots mark the Bragg peak positions. **e,** Disappearance of δZ signal at the critical magnetic field (left) and at the superconducting transition temperature (right). Data are six-fold symmetrized. The black dots mark the Bragg peak positions. **f,** Illustration of inverse magnetic fields applied along Γ-M and Γ-K directions as a time-reversal perturbation of superconductivity. **g, h,** The left panel shows the absence of the δZ signal at the energy outside the superconducting gap, and the right panel shows the emergence of the δZ signal at the energy inside the superconducting gap. Reversed magnetic fields are applied along Γ-M direction (**g**) and Γ-K direction (**h**), respectively. The data are vertically and horizontally symmetrized. All the data were taken at $V$ = 5mV, $I$ = 1nA, $T$ = 0.3K, except for (**e**) right panel taken at $T$ = 5K.

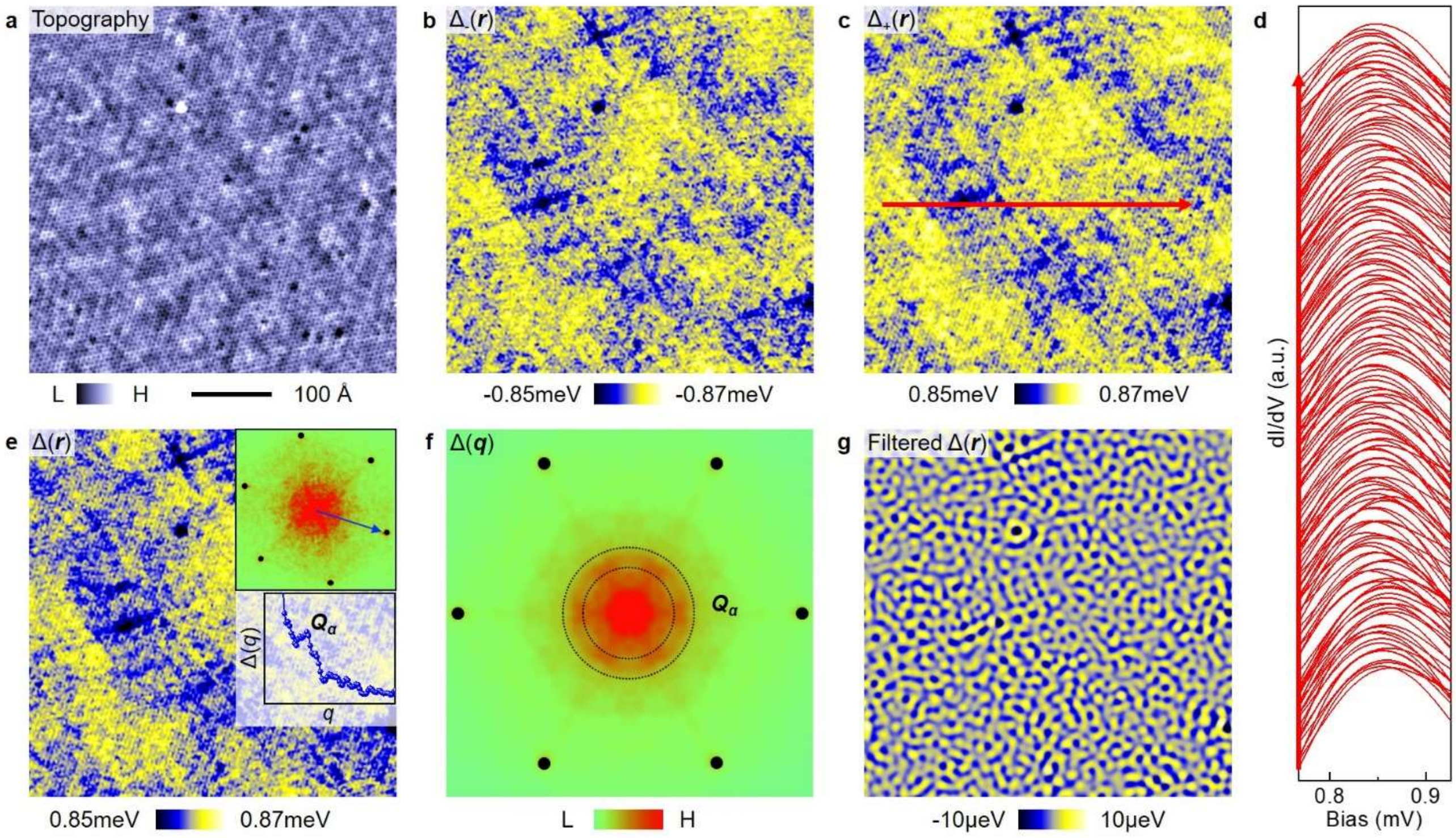

**Figure 3 Spontaneous Cooper pairing modulation. a,** Topographic image for a clean Sb surface, where the following gap map is taken. **b,** Energy map of superconducting coherence peak at negative bias $\Delta_-(\boldsymbol{r})$. **c,** Energy map of superconducting coherence peak at positive bias $\Delta_+(\boldsymbol{r})$. **d,** Spectrums of the positive coherence peak along the line in **c**. From the peak position of each curve, we determine $\Delta_+$ at each location. Spectrums are offset for clarity. **e,** Superconducting gap map $\Delta(\boldsymbol{r}) = [\Delta_+(\boldsymbol{r}) - \Delta_-(\boldsymbol{r})]/2$. The upper inset shows its Fourier transform with an arrow along the Bragg direction, and the lower inset shows the corresponding line cut with a state at $\boldsymbol{Q}_\alpha$. **f,** Six-fold symmetrized Fourier transform of the gap map. The Dash lines mark $\boldsymbol{Q}_\alpha$ in between. **g,** Inverse Fourier transform of $\Delta(\boldsymbol{q})$ for the areas around $\boldsymbol{Q}_\alpha$ in the inset of **e**. All the data were taken at $V = 5$mV, $I = 1$nA, $T = 0.3$K.

**Data availability**
All data are available in the main text or the supplementary materials.

**Competing interests**
The authors declare no competing interests.

## Methods

Single crystal growth

Single crystals of $Cs(V_{0.86}Ta_{0.14})_3Sb_5$ were synthesized via the self-flux method by using $Cs_{0.4}Sb_{0.6}$ as the flux (Cs, bulk, 99.8%; V, piece, 99.999%; Ta, powder, 99.99%; Sb, shot, 99.9999%). The above materials were loaded into an alumina crucible and then heated to 1000 °C for 200 h. After holding for 10 h, the mixture was cooled to 200°C at 260 h at a rate of 3 K/h. When dropped down to room temperature, the furnace was turned off. To remove the flux, the obtained samples were soaked in deionized water. Finally, shiny single crystals with hexagonal features were obtained. Figure. S1 shows the zero-resistivity data, signaling a critical temperature $T_C$ = 5K.

Scanning tunneling microscopy

Single crystals with sizes up to 3mm×3mm×1mm were cleaved mechanically *in situ* at 10K in ultra-high vacuum conditions, and then immediately inserted into the microscope head, already at He4 base temperature (4.2K). We then further cool the microscope head to 0.3K via a He3-based single-shot refrigerator. The magnetic field was applied with a small ramping speed of 1T per 20mins. After ramping the field to a desired value, the superconducting magnet is set in the persistent mode, after which we wait for 1~2h for the system to relax and then find the same atomic position and start to take spectroscopic measurements. Tunneling conductance spectra were obtained with Ir/Pt tips using standard lock-in amplifier techniques with a root mean square oscillation voltage of $V_m$ = 0.05meV under applied bias voltage of V = 5mV and tunneling current I = 1nA. We extensively scan each crystal for large and clean Sb surfaces, which can take up to one week. Topographic images were taken with tunneling junction set up: V = -100~-200mV I = 0.05~0.5nA. The conductance maps and gap map were obtained by taking a spectrum at each location (off feedback loop) with tunneling junction set up: V = 5mV, I = 1nA, and modulation voltage $V_m$=0.05~0.2mV. The tunneling spectrum at 30mK is taken with a separate dilution refrigerator-based scanning tunneling microscope with the same scanning and tunneling setup, except for a modulation voltage of $V_m$ = 0.02mV. The symmetrization process of the data includes six-fold symmetrization and mirror symmetrization.

Muon spin resonance (μSR)

Muon spin resonance (μSR) is a magnetic sensitive probe for internal magnetism of a many-body ordering state. The μSR experiments were carried out at the Swiss Muon Source (SμS) Paul Scherrer Insitute, Villigen, Switzerland. Zero field (ZF) and transverse field (TF) μSR experiments on the single crystalline samples were performed on the high-field HAL-9500 and general purpose surface-muon (GPS) instruments at the Swiss Muon Source (SμS) at the Paul Scherrer Institut, in Villigen,

Switzerland. Zero field is dynamically obtained (compensation better than 30 mG) by a newly installed automatic compensation device. When performing measurements in zero-field the geomagnetic field or any stray fields are tabulated and automatically compensated by the automatic compensation device.

**Author contributions**

H.D., G.L., and T.Y. conducted the scanning tunneling microscopy experiments in consultation with J.X.Y.; Z.G., C.M., R.K. and H.L. conducted the muon spin resonance experiments; J.L. and Z.W. synthesized and characterized the transport of samples; Y.X., W.L., F.B., P.B. and P.D. conducted polarized neutron scattering experiments; S.S., H.M., G.C. and J.L. conducted first-principles calculations; X.W., M.H.C., A.K., B.M.A., W. H. and Y.Y. contributes to the theoretical understandings; X.Y.Y. and Y.Z. contributed to the calibration of the measurement; H.D. and J.X.Y. performed the data analysis and figure development and wrote the paper with contributions from all authors; J.X.Y. supervised the project.

**Correspondence and requests for materials** should be addressed to J.X.Y.

**Acknowledgments**

We thank the insightful discussions with Titus Neupert, Qianghua Wang, Dalila Bounoua and Yvan Sidis. We are also grateful to the full IRIG/D-phy/MEM/MDN group of CEA Grenoble who helped us with the experimental setup of neutron scattering. We acknowledge the support from the National Key R&D Program of China (Nos. 2023YFA1407300, 2023YFF0718403, 2022YFA1403400, 2020YFA0308800), the National Science Foundation of China (Nos. 12374060, 12321004, 12234003), and Guangdong Provincial Quantum Science Strategic Initiative (GDZX2201001). Z.W. also acknowledge the Beijing Natural Science Foundation (Grant No. Z210006), the Beijing National Laboratory for Condensed Matter Physics (Grant No. 2023BNLCMPKF007), and the Analysis and Testing Center at BIT for assistance in facility support. Work at Nanyang Technological University was supported by the National Research Foundation, Singapore, under its Fellowship Award (NRF-NRFF13-2021-0010), the Agency for Science, Technology and Research (A*STAR) under its Manufacturing, Trade and Connectivity (MTC) Individual Research Grant (IRG) (Grant No.: M23M6c0100), Singapore Ministry of Education (MOE) AcRF Tier 2 grant (MOE-T2EP50222-0014) and the Nanyang Assistant Professorship grant (NTU-SUG). P.D. is supported by the U.S. DOE, BES under Grant No. DE-SC0012311. A.K. acknowledges support by the Danish National Committee for Research Infrastructure (NUFI) through the ESS-Lighthouse Q-MAT. Z.G. acknowledges support from the Swiss National Science Foundation (SNSF) through SNSF Starting Grant (No. TMSGI2${_}$211750).

**Supplementary Materials for**
**Evidence chain for time-reversal symmetry-breaking kagome superconductivity**

**Table of Contents**

Figure S20. Zero-field μSR data.
Figure S21. Polarized neutron scattering experiment

Table S1, Spin-flip intensities at the Bragg peak (1,1,1) at 6 K for a monitor M = 1e6 (corresponding to a counting time of 237 sec).
Table S2, Spin-flip intensities at the Bragg peak (1,1,0) at 7 K and 1.5 K for a counting time of 1 minute.
Table S3, Spin-flip intensities at the CDW position (0.5, 0.5, 0) for a monitor M = 1e6 (corresponding to a counting time of 237 sec).

## 1. Single crystal superconducting transition

Figure. S1 shows the zero-resistivity data, signaling a critical temperature $T_C$ = 5K.

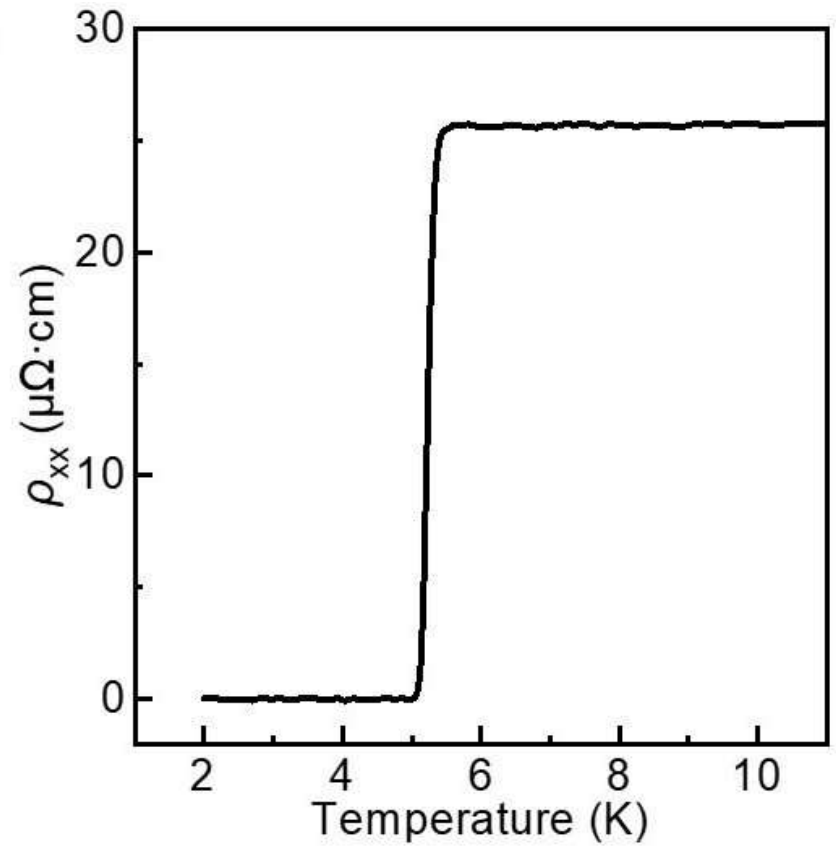

**Figure S1 Zero-resistivity data.** Longitudinal resistivity turns to zero below $T_C$ = 5K.

## 2. Phase diagram and scattering source analysis

Figure S2**a** shows the schematic phase diagram of $Cs(V,Ta)_3Sb_5$. Figure S2**b** and **c** show the atomic dopant imaging at a higher bias voltage and corresponding low bias image. Our counting of these dopants is consistent with the nominal doping. Besides the dopants, there are also Sb vacancies that exist on the surface. Eventually, the scattering source is predominantly Ta dopants and a few Sb vacancies, both of which are expected to be non-magnetic in nature.

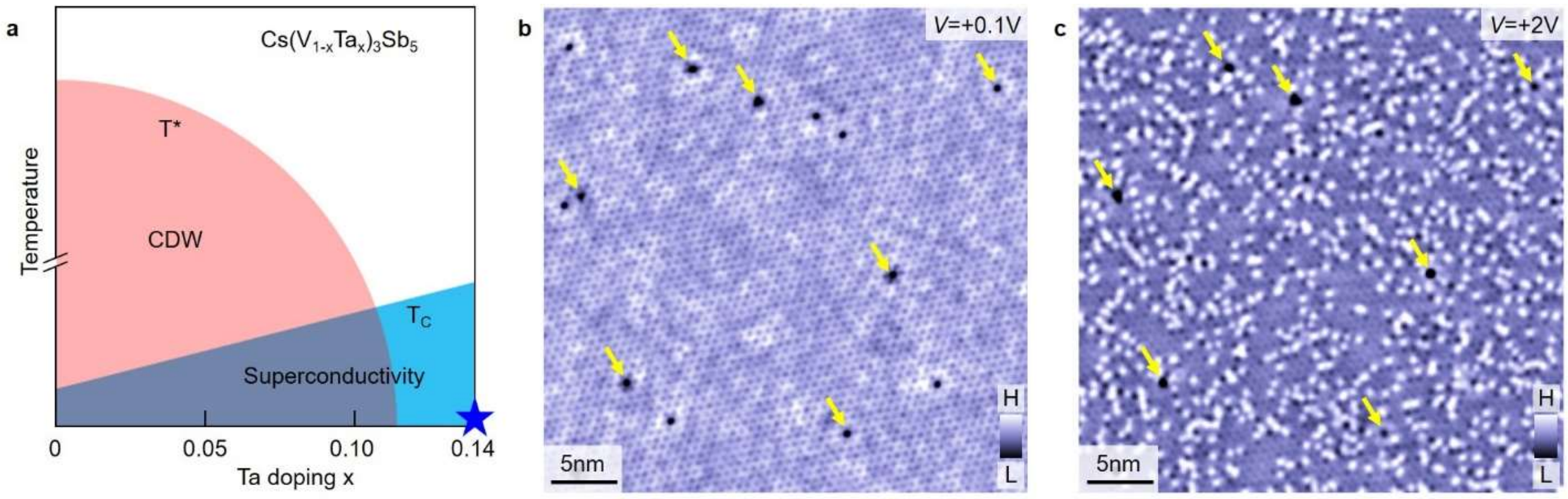

**Figure S2 Atomic dopant identification. a,** Schematic Phase diagram $Cs(V_{1-x}Ta_x)_3Sb_5$. The doping level of our sample is marked by the blue star. **b,** A topographic image of the Sb surface layer with a bias voltage of 0.1V. **c,** A topographic image taken at the same area with a bias voltage of 2V, showing atomic dopants as individual bright spots. The concentration of these dopants is found to be consistent with the nominal doping level ~14%. Yellow arrows mark the same defects shown in the two images.

## 3. Normal state quasi-particle interference and absence of nematicity

Figures S3**a** and **b** show the normal state differential conductance mapping taken at $E_F$ at 5K, the Fourier transform of which gives the quasi-particle interference. Since we do not detect substantial nematicity or chirality, we then perform six-fold symmetrization for our interference data as used in our main paper. The lack of nematicity can be due to the full suppression of charge density wave order

existing in the parent compound as discussed previously[45-48]. Figure S3**c** shows the interference signals as a function of energy along two high-symmetry directions. The low-energy quasi-particle interference signals are highly dispersive, thus are directly associated with electronic band dispersions.

It is also worthwhile to mention that our measurement shows a full gap state, in comparison with the existence of finite states within the superconducting gap in the charge ordered $CsV_3Sb_5$ by previous tunneling studies[27,49,50], despite that transport and muon spin resonance evidence for a full gap pairing state[14,51]. The existence of residual states in the tunneling study of $CsV_3Sb_5$ is still unclear, likely owing to the pair density wave order[52,53] or the interplay with the time-reversal symmetry-breaking charge order[8-15,54-57].

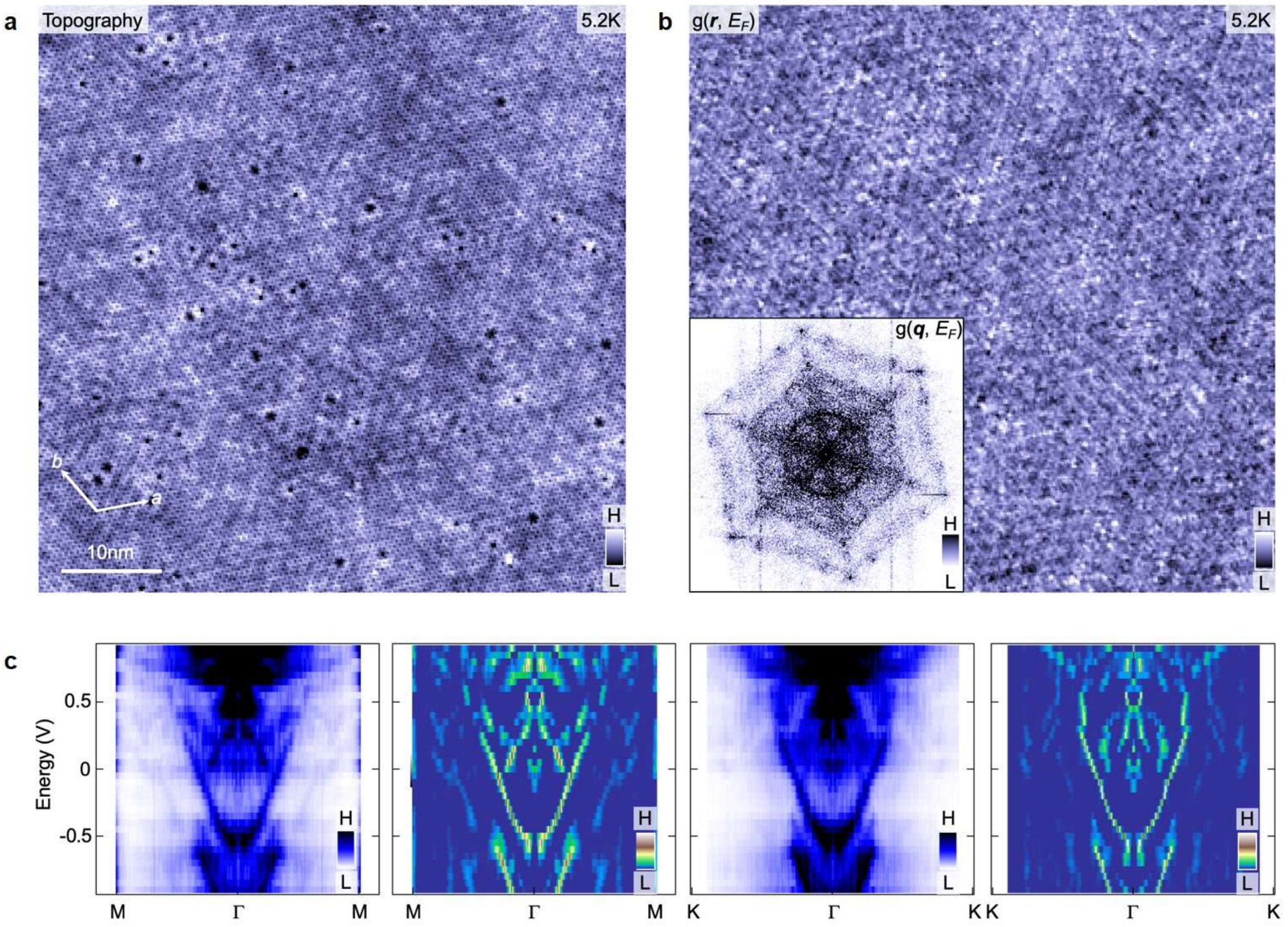

**Figure S3 Normal state quasi-particle interference and dispersion. a,** Atomic resolved topographic image of Sb surface. **b,** Differential conductance map taken at the Fermi-level at 5.2K. The inset shows its Fourier transformation (unsymmetrized). **c,** QPI dispersion along two high symmetry directions (Γ-M and Γ-K) as well as their corresponding curvature analysis shown on the right.

## 4. Magnetic field enhanced Bogoliubov quasi-particle interference

Figure S4 shows more data and analysis in further supporting our claim in Figure. 2**b**. In Figs. S4**a-c**, we show the correspondence between real space data and q-space data. Figure S4**d** shows the differential conductance spectrum taken in between vortices. The superconducting gap disappears at 1.5T, which defines the critical magnetic field for the disappearance of the superconducting signal. Figure. S4**e** further shows the field-induced enhancement of the intensity for all three Q vectors.

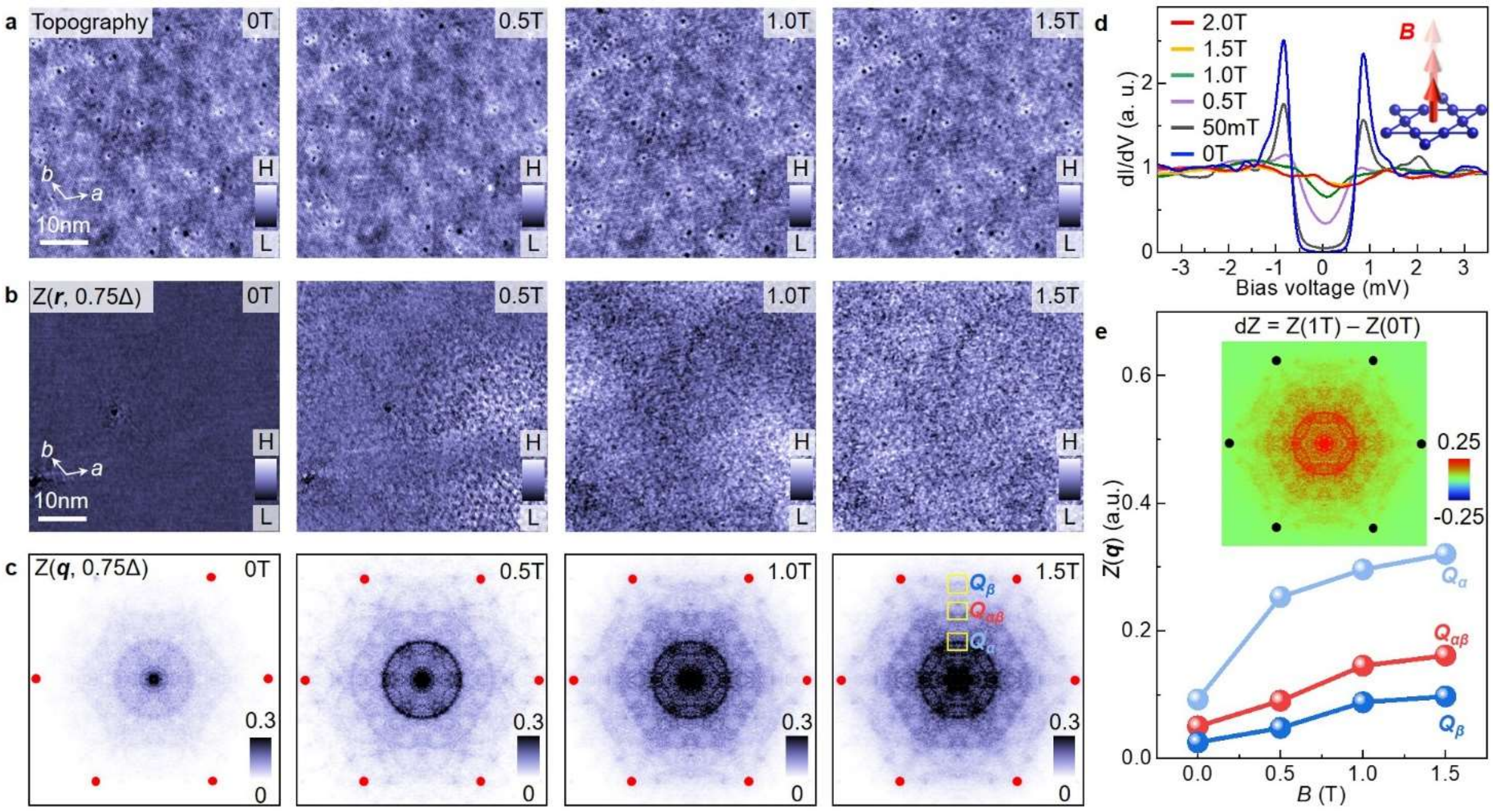

**Figure S4 Magnetic field enhanced Bogoliubov quasi-particle interference. a,** Topographic images of the same large field of view under different magnetic fields applied perpendicular to the kagome lattice. **b,** Corresponding Z maps at different fields, where Z($r$, $E$) = g($r$, +$E$)/g($r$, -$E$). **c,** Corresponding Fourier transform of Z($r$) as Bogoliubov quasi-particle interference $Z(q)$ at different fields. The intensities at $Q_\alpha$, $Q_\beta$ and $Q_{\alpha\beta}$ are all progressively increasing with increasing field strength. Data are six-fold symmetrized. The six red dots mark the expected Bragg spots. **d,** Differential conductance spectrums measured at different magnetic fields applied perpendicular to the kagome lattice (as illustrated in the inset). The saturating field is around 1.5T, which denotes the spectroscopically defined critical magnetic field. **e,** Magnetic field evolution of Bogoliubov QPI Z($q$). Inset plots the difference of Z($q$) between 1T and 0T. All the data are taken at the same atomic position at 0.3K.

## 5. Experimental process to obtain the δZ(*q*) data

In Fig. S5, we show key steps to obtain the δZ($\boldsymbol{q}$) data. We first take differential conductance maps g($E$,$\boldsymbol{r}$) and obtain the Z($E$,$\boldsymbol{r}$)=g(+$E$,$\boldsymbol{r}$)/g(-$E$,$\boldsymbol{r}$) maps (Z map selects Bogoliubov quasi-particles with particle-hole symmetry) for the same field of view at three representative energies $E$=0.25Δ, $E$=0.75Δ, $E$=1.25Δ. A pair of Z maps are taken for $B$=0.5T and $B$=-0.5T. Taking into account the preparation time for cooling to 300mK and relaxation time for each field ramping, each pair of Z maps would take nearly a week. Then, we perform the Fourier transform for each Z($E$, $\boldsymbol{r}$) map to obtain Z($E$, $\boldsymbol{q}$). Note that Z($E$, $\boldsymbol{q}$) maps are symmetrized for the enhancement of signal-to-noise ratio. Lastly, we obtain the δZ($\boldsymbol{q}$) data as δZ(E, $\boldsymbol{q}$)= δZ(E,$\boldsymbol{q}$,+$B$)- δZ(E,$\boldsymbol{q}$,-$B$) for these three representative energies. We further present the unsymmetrized δZ($\boldsymbol{q}$) data to support that the key information is independent of the symmetrization process.

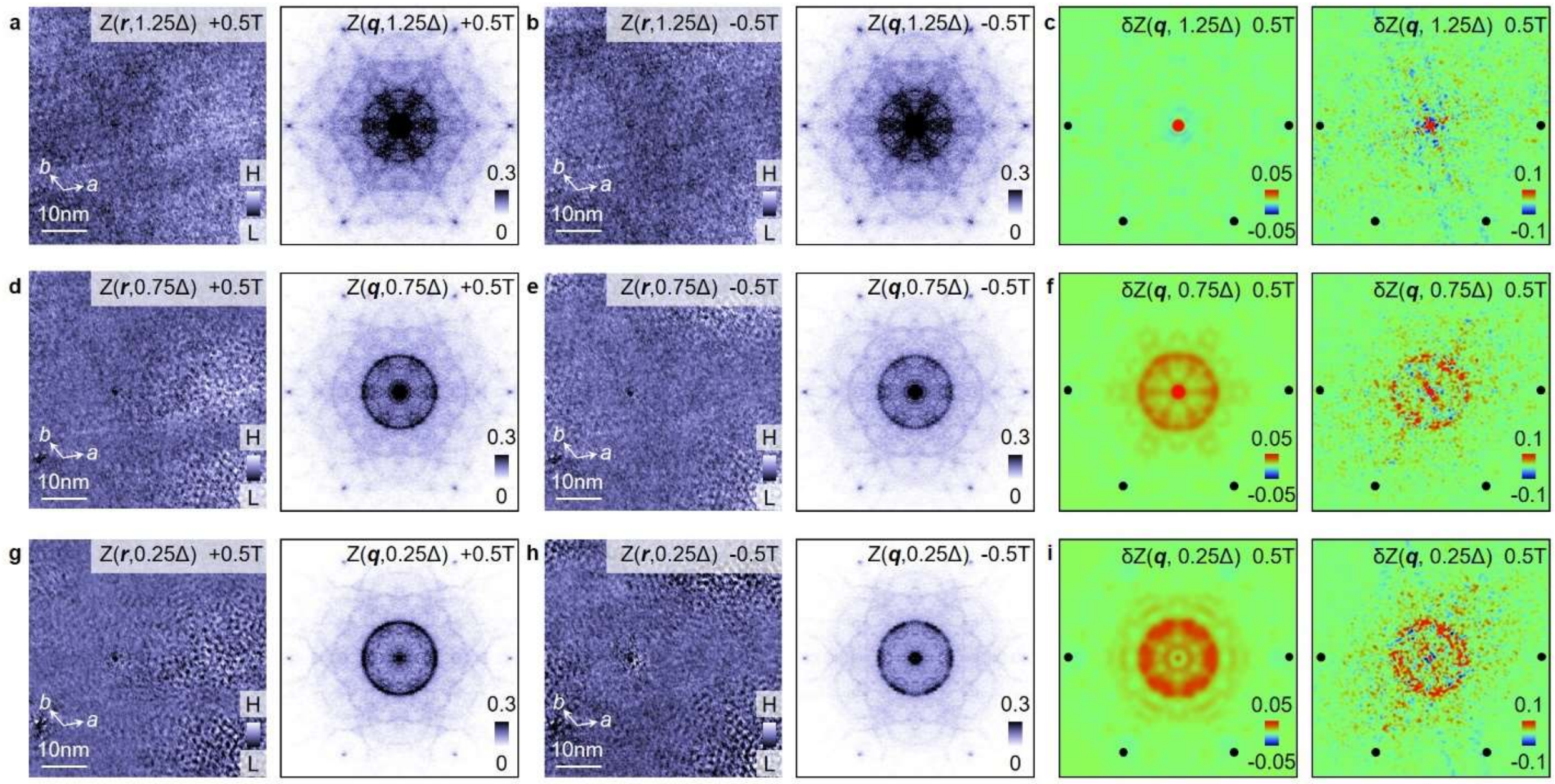

**Figure S5 Process to obtain the δZ(*q*) data. a, d, g,** Z(***r***) map and corresponding Fourier transform Z(***q***) taken at *B*=+0.5T for energy *E* = 1.25Δ, 0.75Δ, and 0.25Δ, respectively. **b, e, h,** Z(***r***) map and corresponding Fourier transform Z(***q***) taken at *B*=-0.5T for energy *E* = 1.25Δ, 0.75Δ, and 0.25Δ, respectively. **c, f, I,** δZ(***q***) data for energy *E* = 0.25Δ, 0.75Δ, and 1.25Δ, respectively. The left data is six-fold symmetrized, and the right data is the unsymmetrized.

## 6. δZ(*q*) data under different fields and temperatures

Figure S6 shows both the field dependence and temperature dependence of δZ(***q***) data (unsymmetrized) in further supporting our claim in Fig. 2**e**. The δZ(***q***) signal substantially weakens as we increase the field or temperature to destroy superconductivity, indicating an intimate relationship between the TRSB quasi-particle interference signal and superconductivity.

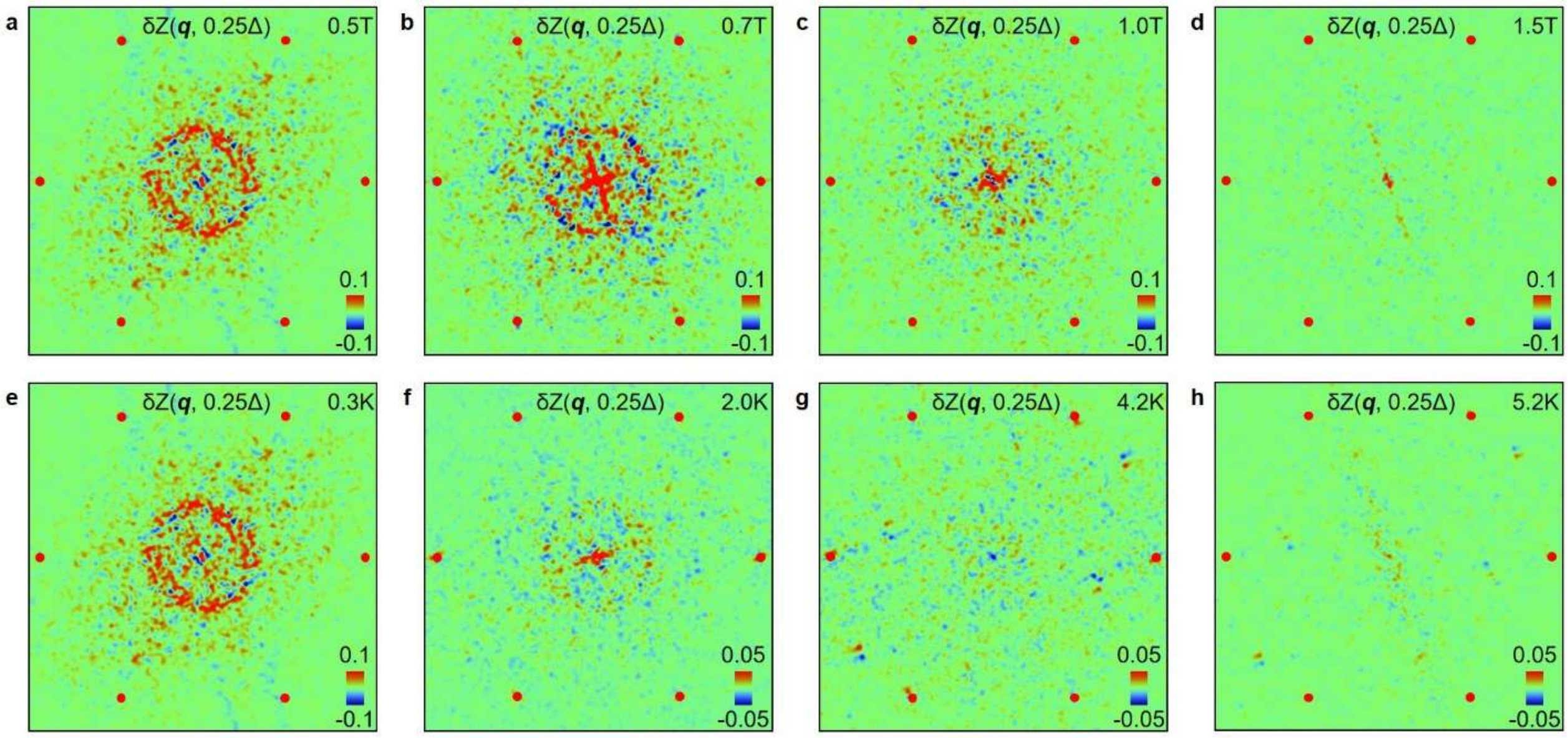

**Figure S6 δZ(*q*) data under different fields and temperatures. a-d,** Field dependent δZ(*q*) data at 0.3K. The magnetic field is applied along c-axis. **e-h,** Temperature dependent δZ(*q*) data at *B* = ±0.5T

applied along c-axis. Data are not symmetrized.

## 7. Robust time-reversal asymmetrical quasi-particle interference in δg(*q*) data

In the main text, we use δZ(***q***) data to elucidate the electronic signature of TRSB, as the Z map is known to select Bogoliubov quasi-particles. In Fig. S7, we show that the key observations from δZ(***q***) data, including its appearance mainly at $\boldsymbol{q} = \boldsymbol{Q}_\alpha$ and disappearance at energy outside the pairing gap, at critical field, and at $T_C$, can all be reasonably reproduced by the δg(***q***) data, suggesting the robust phenomena of TRSB, which is independent of our data processing methods.

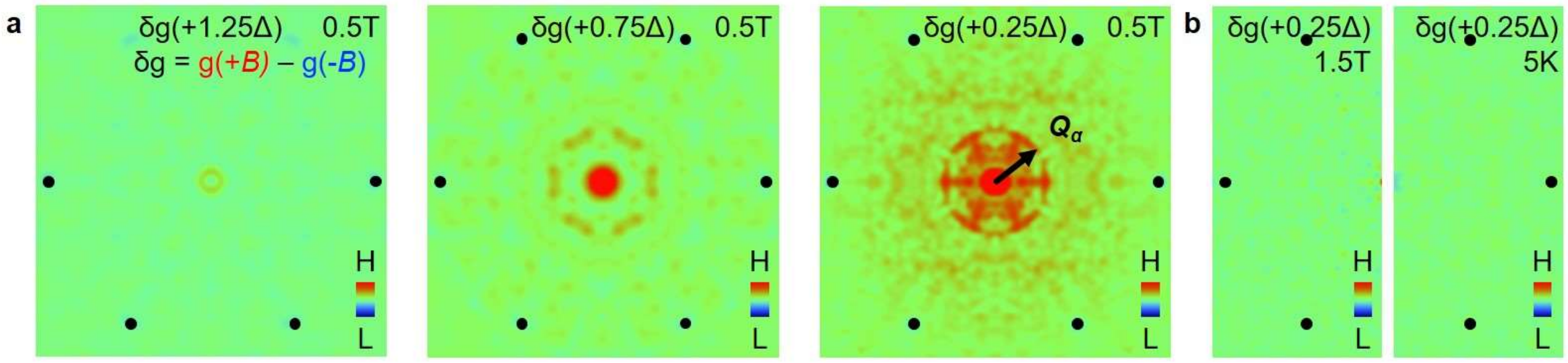

**Figure S7 Robust time-reversal asymmetrical quasi-particle interference in δg(*q*) data. a,** Similar to the δZ(***q***) data in Fig. 2**d** and Fig. S5, we process δg(***q***) data for B=±0.5T for three representative energies. Note that Z(*E*,***r***)=g(+*E*,***r***)/g(-*E*,***r***). **b,** Disappearance of δg data at critical field and $T_C$.

## 8. Ruling out trivial reasons for δZ(*q*) signal

Electronic signal for a TRSB superconducting state has long been elusive for tunneling experiments. Based on the TRSB-sensitive setup using a pair of reverse magnetic fields, in the main text we show that δZ(***q***) signal δZ(***q***) = Z(***q***, +*B*)-Z(***q***, -*B*) can reflect the electronic signature of TRSB superconductivity. Although our observed δZ(***q***) signal is highly expected based on the observation of internal magnetism by muon spin resonance and is highly consistent with Δ(***q***) modulations on the ***q*** vector, to supplement this new technique as a TRSB probe, we have carefully designed several additional experiments to rule out other possible trivial explanations of our data. We now discuss these experiments and our reasoning.

We repeat the field circle and find that the polarity of δZ is not magnetic field history dependent. Namely, we apply +0.5T and perform the Z map; then we apply -0.5T and perform the Z map, and we find δZ(***q***) = Z(***q***, +*B*)-Z(***q***, -*B*) to be positive sign; then we reapply +0.5T and perform the Z map and find δZ(***q***) = Z(***q***, +*B*)-Z(***q***, -*B*) remains to be positive sign. This allows us to rule out the effect of residual vortex pinning owing to external field history on determining our δZ signal. Owing to the small area that our technique can probe, we have yet to observe different domains for δZ with opposite signs for a single sample. However, we have encountered a negative sign of δZ signal for a different sample, suggesting the direction of internal magnetism can be different for different sample areas.

We rule out the trivial reason that the δZ signal is caused by unbalanced ±B field amplitudes (owing to instrument error for example) or substantially different vortex states/numbers within the measured area with opposite fields. These two scenarios can be classified as extrinsic external field effects rather than intrinsic internal field effects. We have two experimental facts that go strongly against the external field effects. Firstly, δZ data disappears for $E$=1.25Δ (this disappearance is expected for internal magnetism of superconductivity) in Fig. 2D left panel. However, as we show in Fig. S8, the Z($\boldsymbol{q}$) signal at E=1.25Δ is very sensitive to an external magnetic field, and its intensity increases with increasing external field, thus unbalanced external field strength will lead to a substantial signal for δZ($\boldsymbol{q}$, 1.25Δ), inconsistent with our observation of the disappearance of δZ($\boldsymbol{q}$, 1.25Δ) signal in Fig. 2**d** left panel. Secondly, the external field affects all three $\boldsymbol{q}$ vectors as demonstrated in Fig. 2**b** and Fig. S4. To show this more clearly, we show the dZ data induced by the external field where dZ($\boldsymbol{q}$) = Z($\boldsymbol{q}$, 0.7T) – Z($\boldsymbol{q}$,0.5T) in Fig. S9, which is substantially different from δZ data. The former shows intensity for all three $q$ vectors as a result of external field effects, while the latter mainly shows intensity for $\boldsymbol{Q}_\alpha$ (again consistent with the Δ($\boldsymbol{q}$) data on $\boldsymbol{q}$ vector).

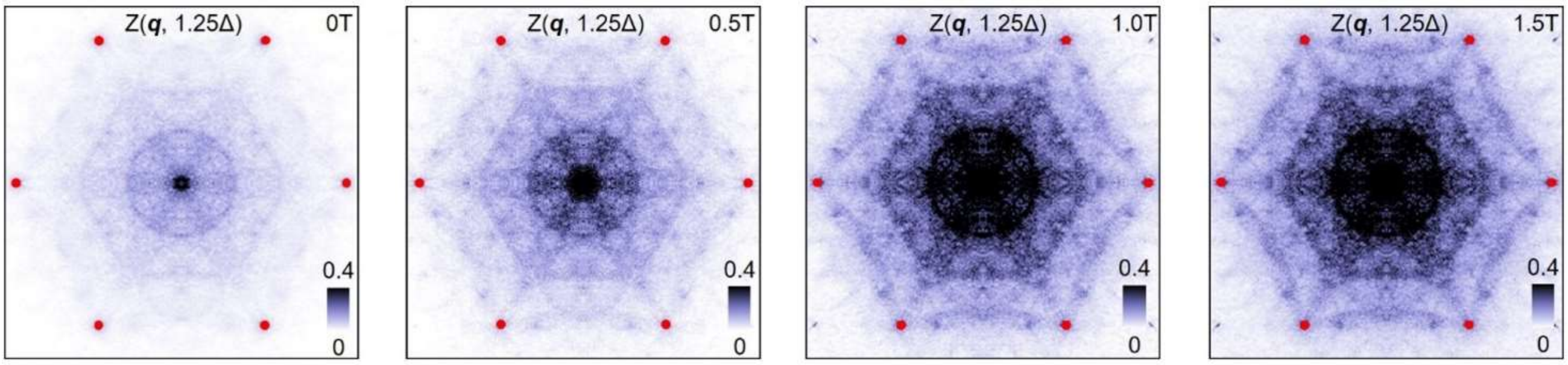

**Figure S8 Field-enhanced quasi-particle interference for $E$ = 1.25Δ.** Z($q$) data taken at $B$ = 0T, 0.5T, 1.0T, and 1.5T. All the data are six-fold symmetrized to enhance the signal-to-noise ratio. All the data are taken at the same atomic position at 0.3K.

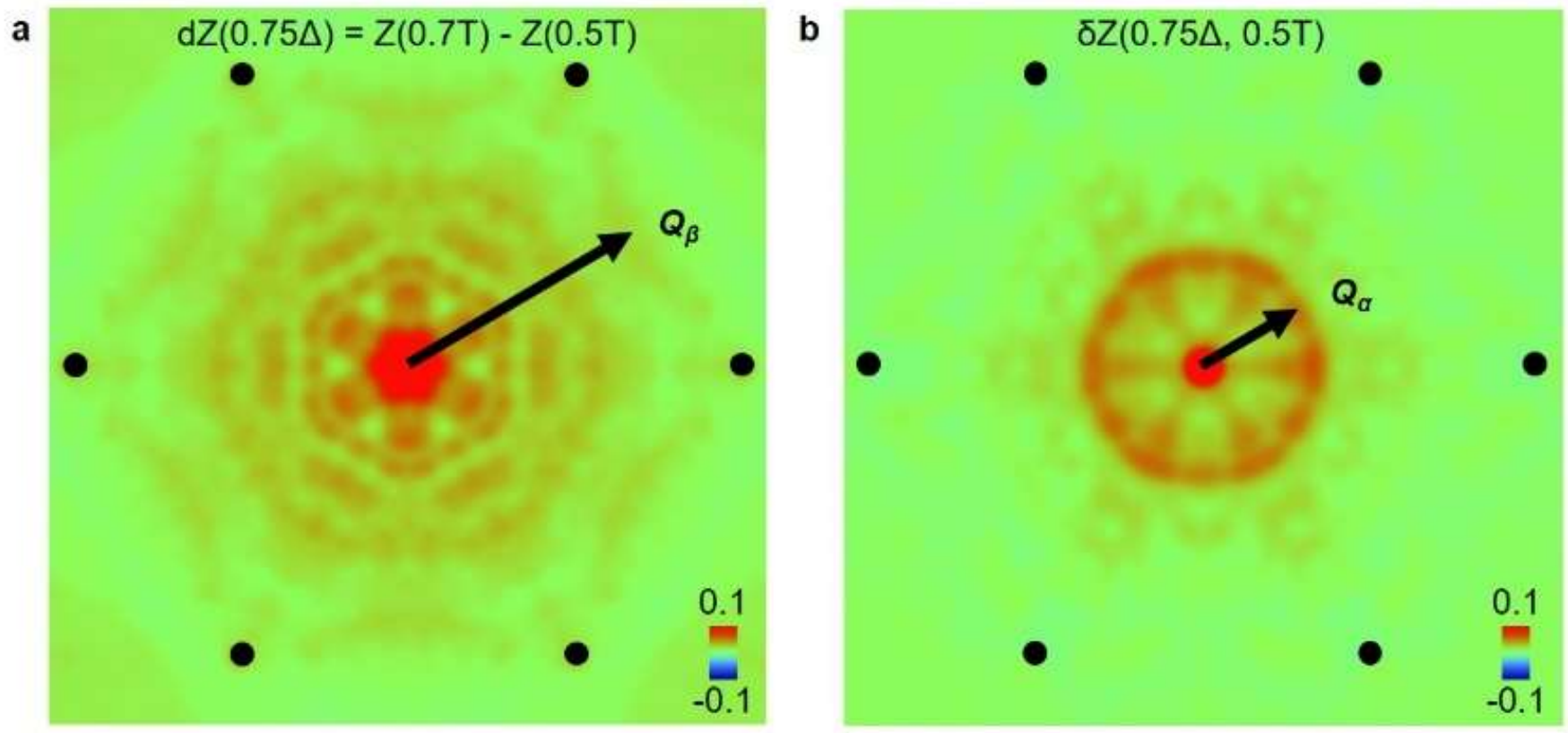

**Figure S9 Comparison between external field (a) and internal field (b) effects on quasi-particle interference.** All the data are six-fold symmetrized to enhance the signal-to-noise ratio. All the data are taken at the same atomic position at 0.3K.

## 9. Vortex matter

We further discuss the impact of vortex matter on the quasi-particle interference. We apply a magnetic field of B=0.5T along the c-axis and measure the tunneling conductance map at zero energy over a large area, which shows a perfect hexagonal vortex lattice (Fig. S10**a**). At a lower applied magnetic field where the vortex is more separated, we can clearly resolve a zero-energy vortex core state (Fig. S10**b**). Then, in an example δZ map data set, we show that its corresponding g map clearly resolves the vortex states (Fig. S11**b**). In the quasi-particle interference Z map, we choose two areas denoted as "off vortex" and "on vortex" and perform Fourier transform analysis in Fig. S11**d-f**. It is shown that they have comparable enhanced signals at $\boldsymbol{Q}_\alpha$. Physically, from the muon spin rotation data in Fig. 1**f** we determine that the magnetic penetration depth is 270nm, while the intervortex distance is typically within 50nm (B from 0.5T to 1T). So at the magnetic field range that we applied in tunneling experiments, the magnetic field is rather homogeneous across the whole area. Therefore, both the on-vortex area and off vortex area contribute to the field-enhanced quasi-particle interference signal.

Since the whole area contributes to the field-enhanced quasi-particle interference, it is fairer to compare the integrated density of states (integrated dI/dV data for the whole area) for a pair of reversed fields such as ±0.5T, as shown in Fig. S12. It is shown that their deviation is below the order of 0.1%, while the δZ/Z is 12%. Therefore, we evaluate that our measured area is sufficiently large to rule out the number fluctuation of vortex states.

To further demonstrate that the region beyond vortex core center also contributes to the time-reversal asymmetrical quasi-particle interference signal, we have performed experiments with applying reversed in-plane fields in Figs. 2**f-h**. Because each vortex contains quantized flux, when the vortex line is parallel to the surface, its center has to be away from the surface. We further support the corresponding unsymmetrized data in Fig. S13.

We further confirm the emergence of δZ($\boldsymbol{q}$=$\boldsymbol{Q}_\alpha$) in the quasi-particle interference map over an extended area (1500Å×1500Å) with four full vortices under reversed magnetic fields of B=±0.5T at our base temperature of T=300mK. First, we use our tip probe to clean out such a clean area (red box in Fig. S14**d**) by pushing the surface Cs adatoms to the surrounding area with tunneling condition V=5mV and I=3nA scanning for 24 hours. We then perform the Z map at E=0.25Δ for B=+0.5T and B=-0.5T, respectively. Figures S14**a** and **b** show the Z map and their corresponding Fourier transforms Z($\boldsymbol{q}$) under B=±0.5T, respectively. Owing to the limited pixel (256×256, and a larger pixel map will take longer time than our holding time of three days at 300mK), the Z($\boldsymbol{q}$) data could not cover the original Bragg peaks but be enough to cover $\boldsymbol{Q}_\alpha$. The six vector peaks are the folded Bragg peaks owing to the finite pixels. Crucially, the δZ($\boldsymbol{q}$) map in Fig. S14**c** also detects differential signals at $\boldsymbol{Q}_\alpha$, consistent with our existing conclusion that the TRSB signal in the electronic structure occurs at this scattering channel. We note that δZ/Z = 10±4%.

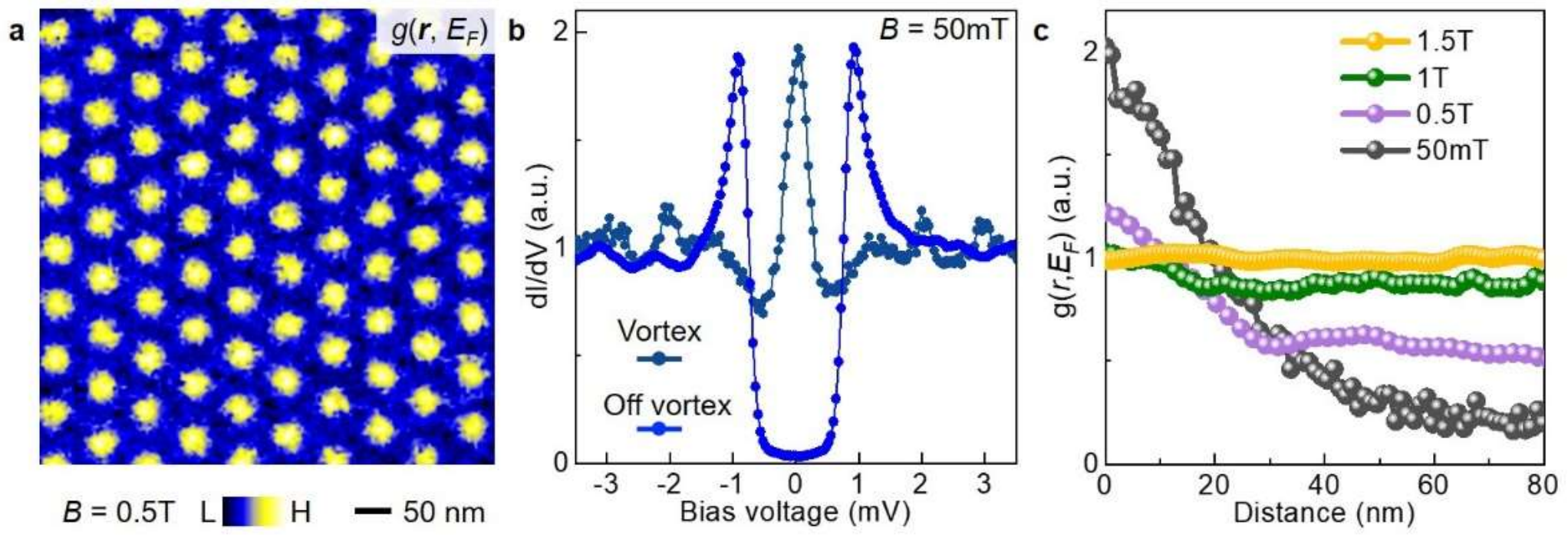

**Figure S10 Vortex lattice imaging (a), vortex core state (b) and linecuts of dI/dV spectrum from core center to far away positions under different magnetic fields (c). At the field of 1.5T, we do not detect clear vortex core states.**

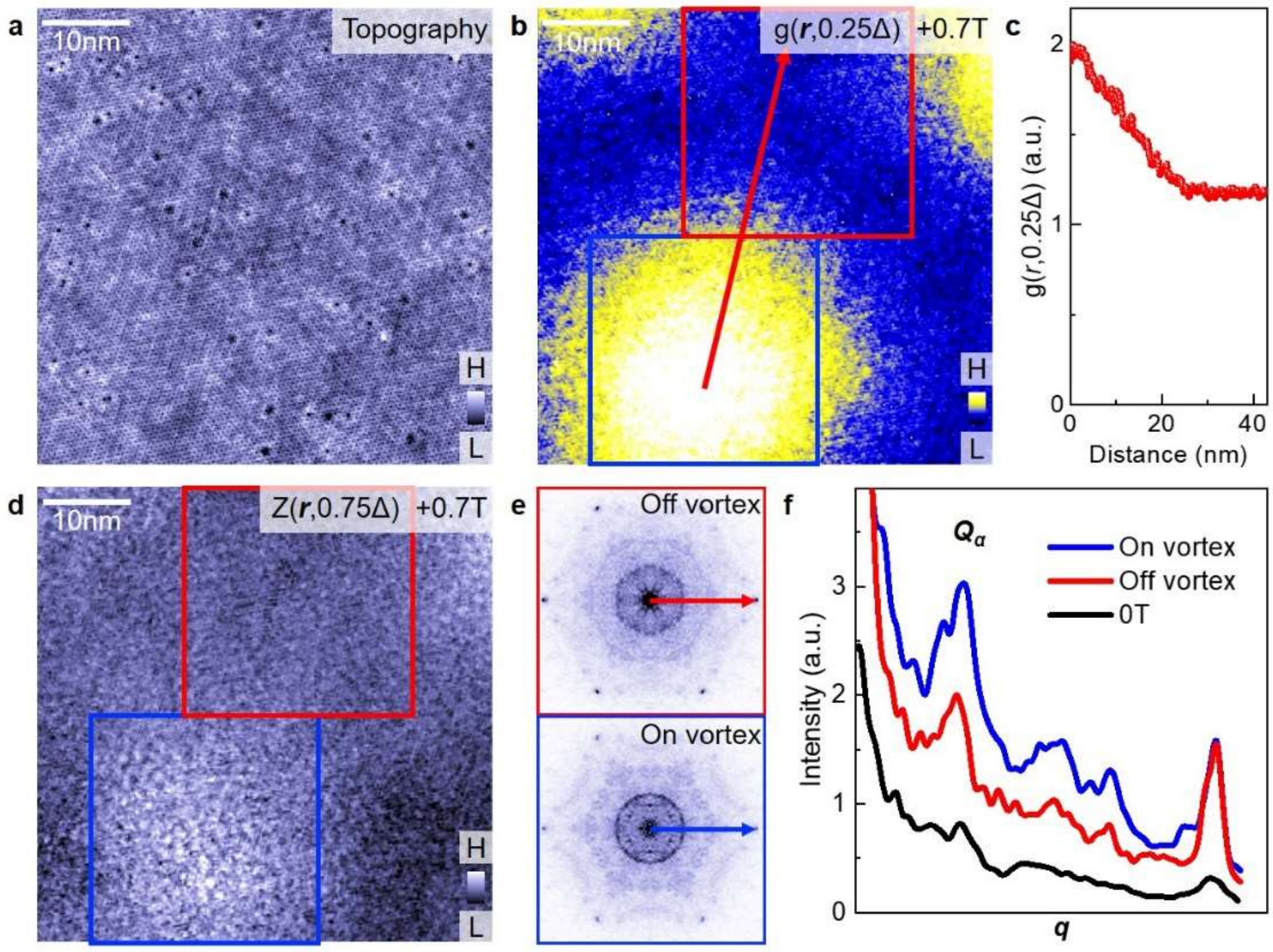

**Figure S11 Enhanced quasi-particle interference from both areas on and off the vortex. a,** Topographic image. **b,** Simultaneously obtained g($\boldsymbol{r}$, 0.25Δ) map showing the vortex core states. **c,** A linecut of the g map from vortex core center to faraway position along the red line shown in **b**. **d,** Simultaneously obtained Z($\boldsymbol{r}$, 0.25Δ) map showing the vortex core states. **e,** Fourier transform of the Z map for the off-vortex area (within the red box in **b** and **d**) and on-vortex area (within the blue box in **b** and **d**). **f,** Comparison of the quasi-particle interference signal for the on-vortex area and off vortex-area, as well as the signal at B=0T. The linecuts are along the lines marked in **e**, showing that they have comparable enhanced signals at $\boldsymbol{Q_\alpha}$.

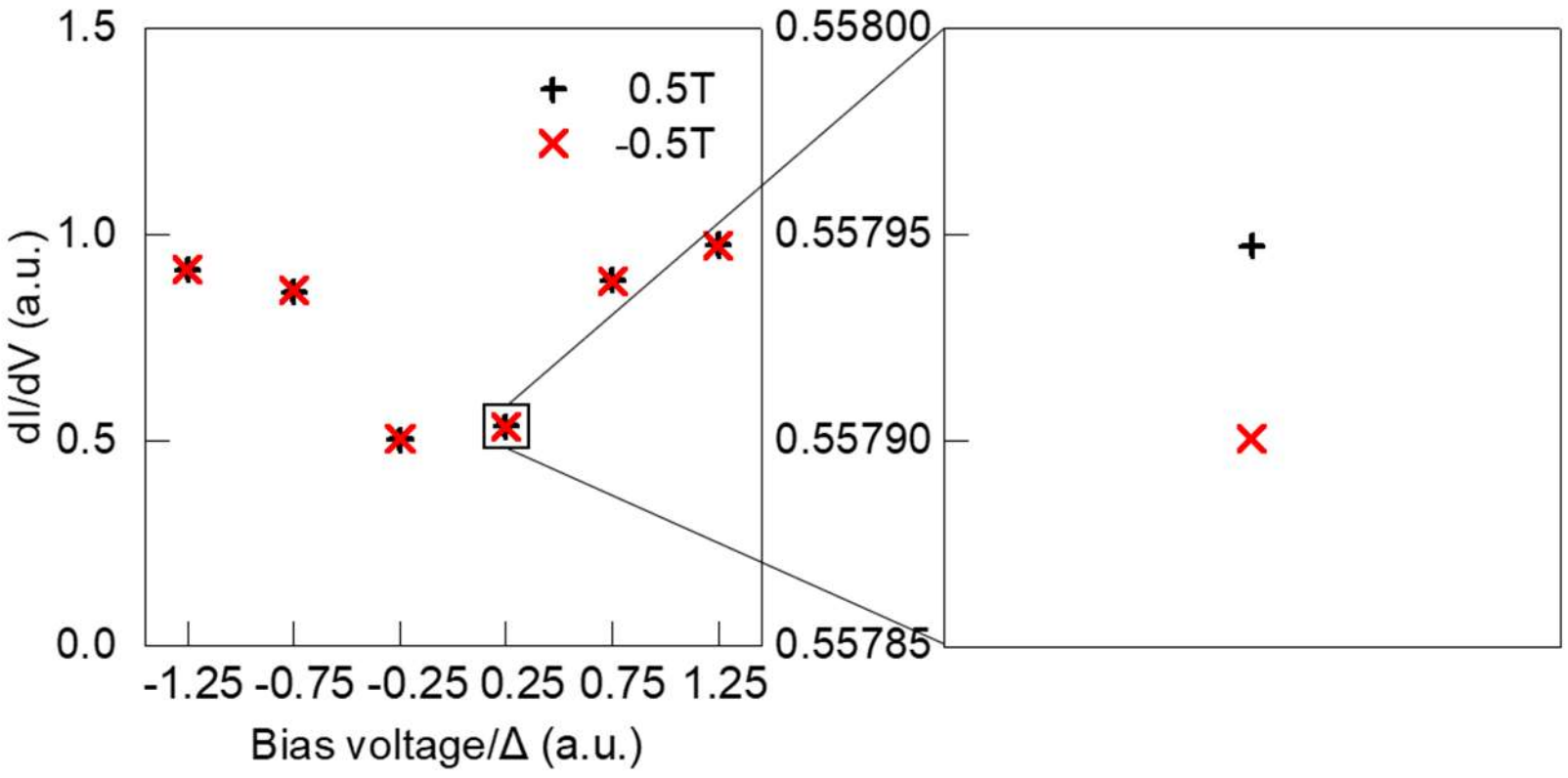

**Figure S12 Integrated dI/dV data for the whole area for ±0.5T.**

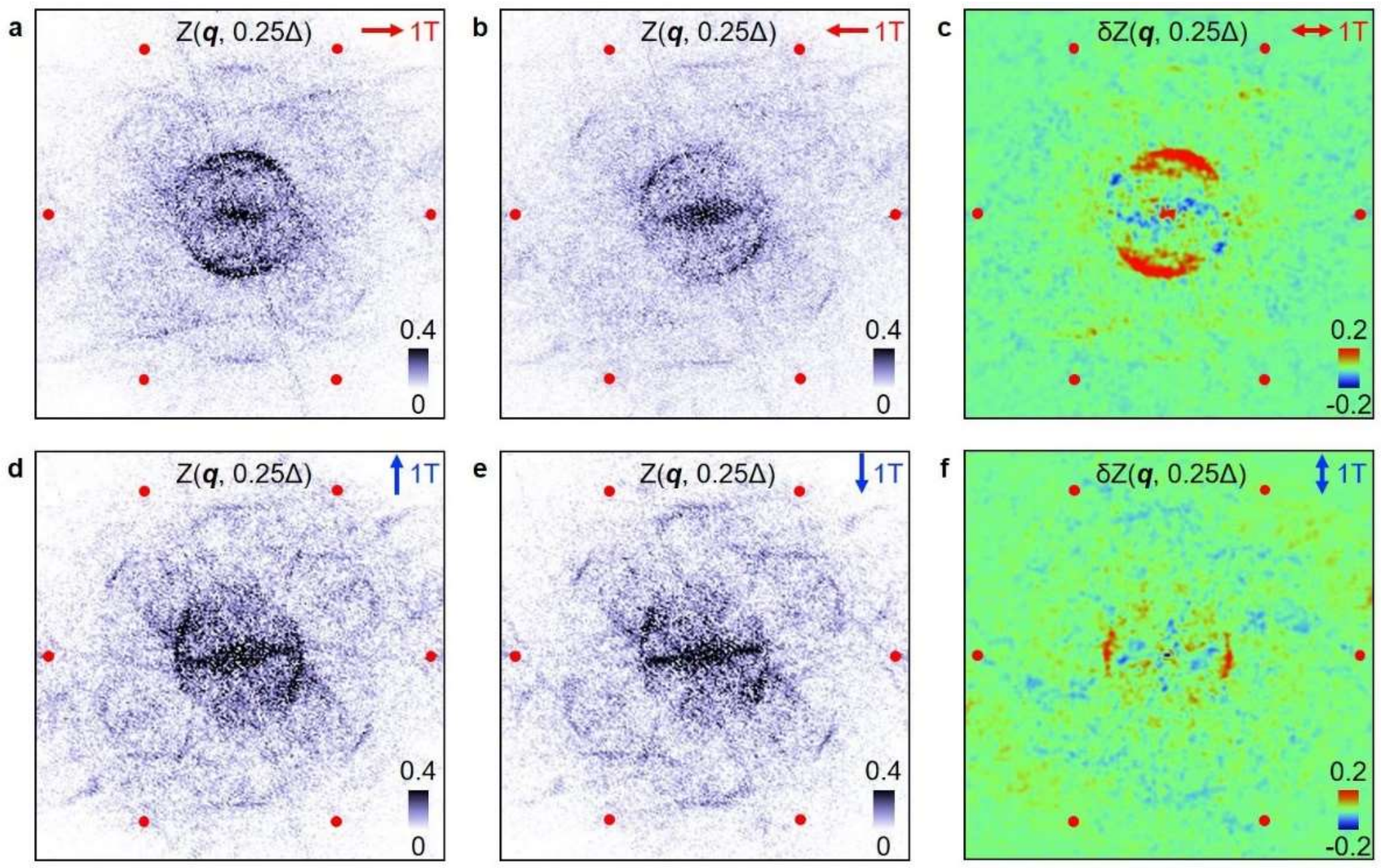

**Figure S13 Unsymmetrized Z(*q*) data (a,b,d,e)and δZ(*q*) data (c,f) for reversed in-plane fields.** The arrows in the up right corners of each figure mark the field direction.

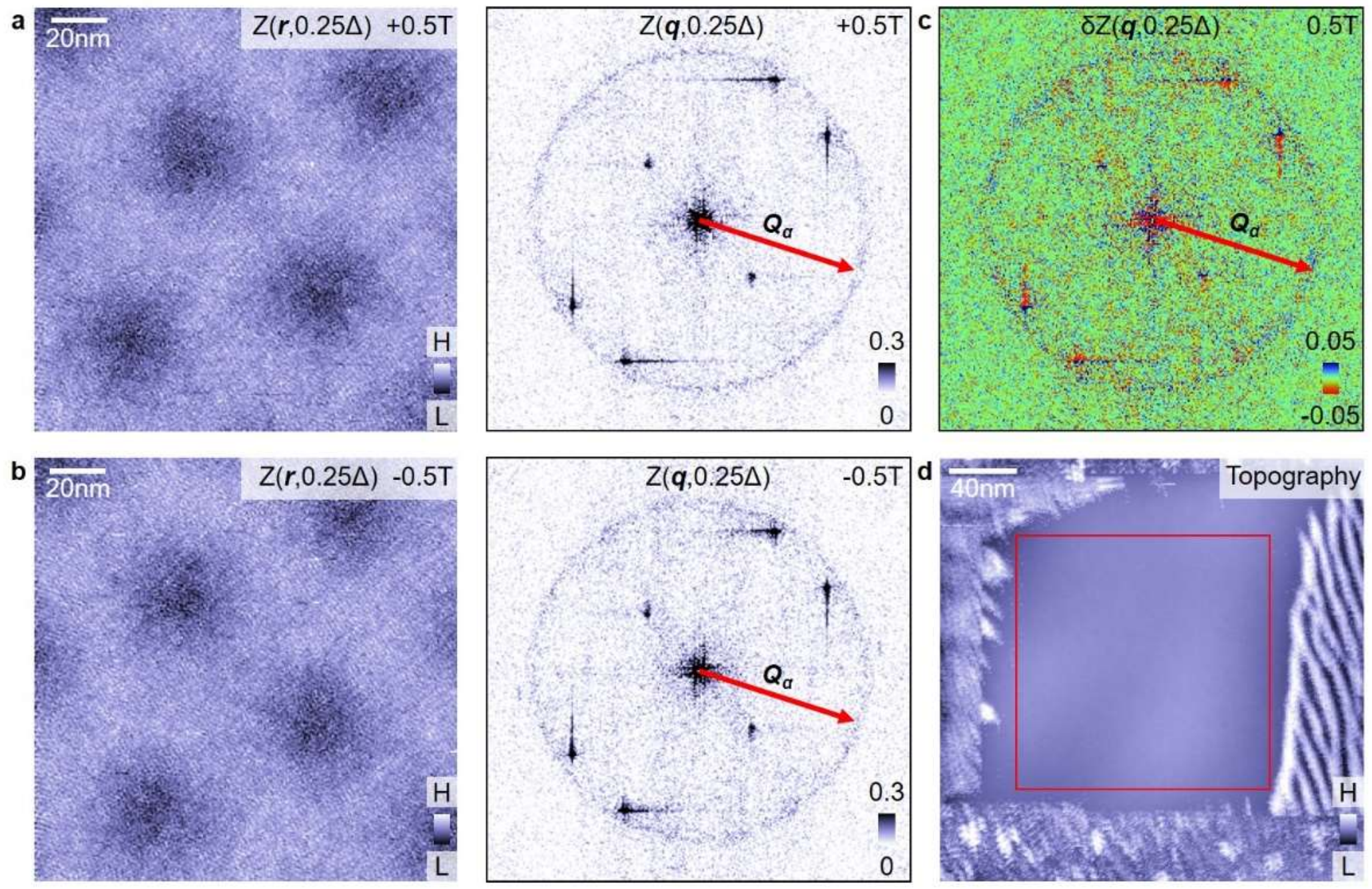

**Figure S14 Reversed field experiment over a large area. a,** Z(*r*) map (left) and corresponding Z(***q***) map (right) for a large area containing four full vortices under a magnetic field of +0.5T applied perpendicular to the lattice. The six peaks are the folded Bragg peaks owing to finite pixels (256×256) over an area of 1500Å×1500Å. **b,** Z(***r***) map (left) and corresponding Z(***q***) map (right) for the same large area under a magnetic field of -0.5T applied perpendicular to the lattice. **c,** δZ(***q***) map for the same area confirming the emergence of TRSB signal at $\boldsymbol{Q}_\alpha$, where δZ(***q***) = Z(***q***, +*B*)-Z(***q***, -*B*). **d,** Topographic image after we clean out an area of 1500Å×1500Å using the tip probe to push the Cs adatoms to surrounding areas. The red box marks the clean area for the Z map. Data are taken at 300mK.

## 10. Gap distribution of the gap map

In Fig. S15 we show the gap distribution of the gap map. We have also checked that the region that we measure gap map also exhibits time-reversal asymmetrical quasi-particle interference (Fig. S16). We have also checked the robustness of the pairing modulation by three different maps: 1) map the coherence peak modulation at $\Delta_+$, taken with a positive bias voltage setup; 2) map the coherence peak modulation at $\Delta_-$, taken with a negative bias voltage setup; 3) map of both coherence peaks to calculate the energy gap as $\Delta=\Delta_+-\Delta_-$, taken with a positive bias voltage setup. The Fourier transforms of these three maps yield the same results that the pair gap modulation occurs at $\boldsymbol{Q}_\alpha$ (Fig. S17).

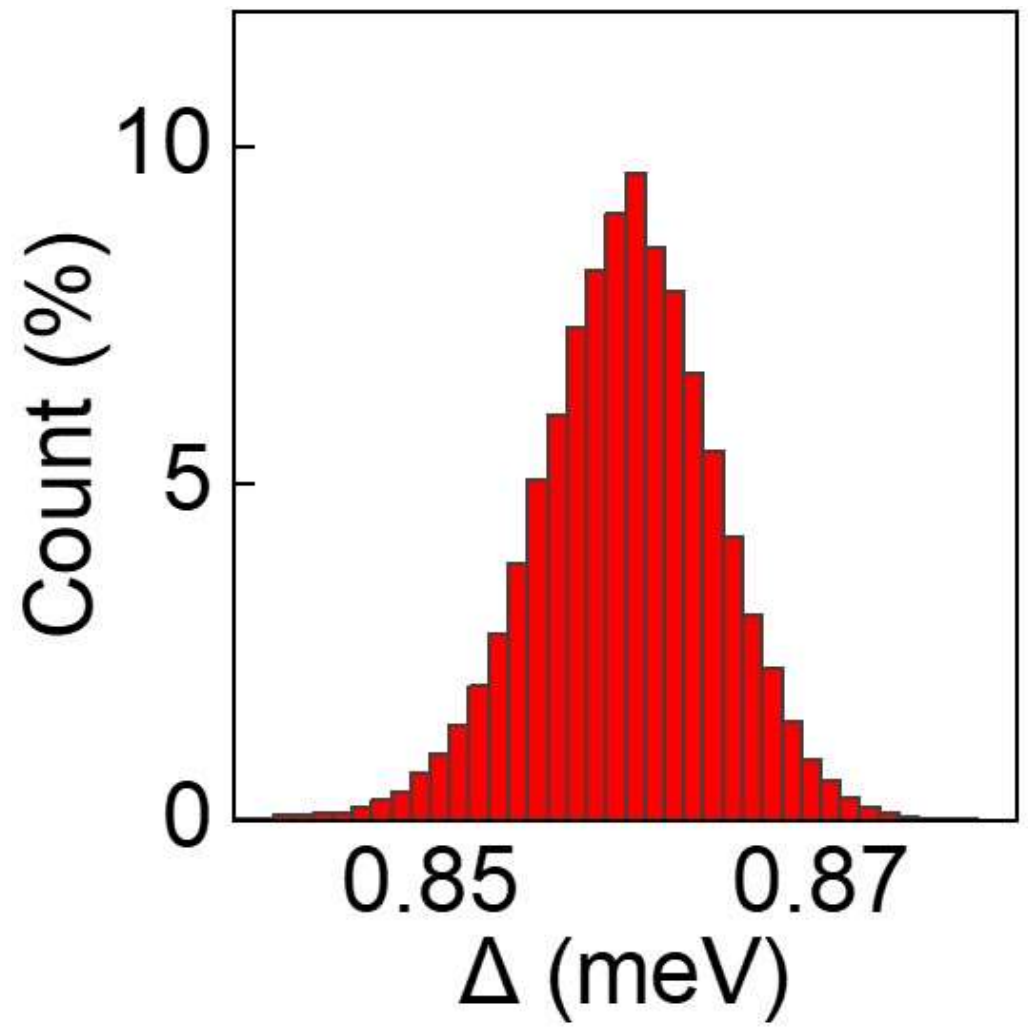

**Figure S15 Gap distribution of the gap map from Fig 3e.**

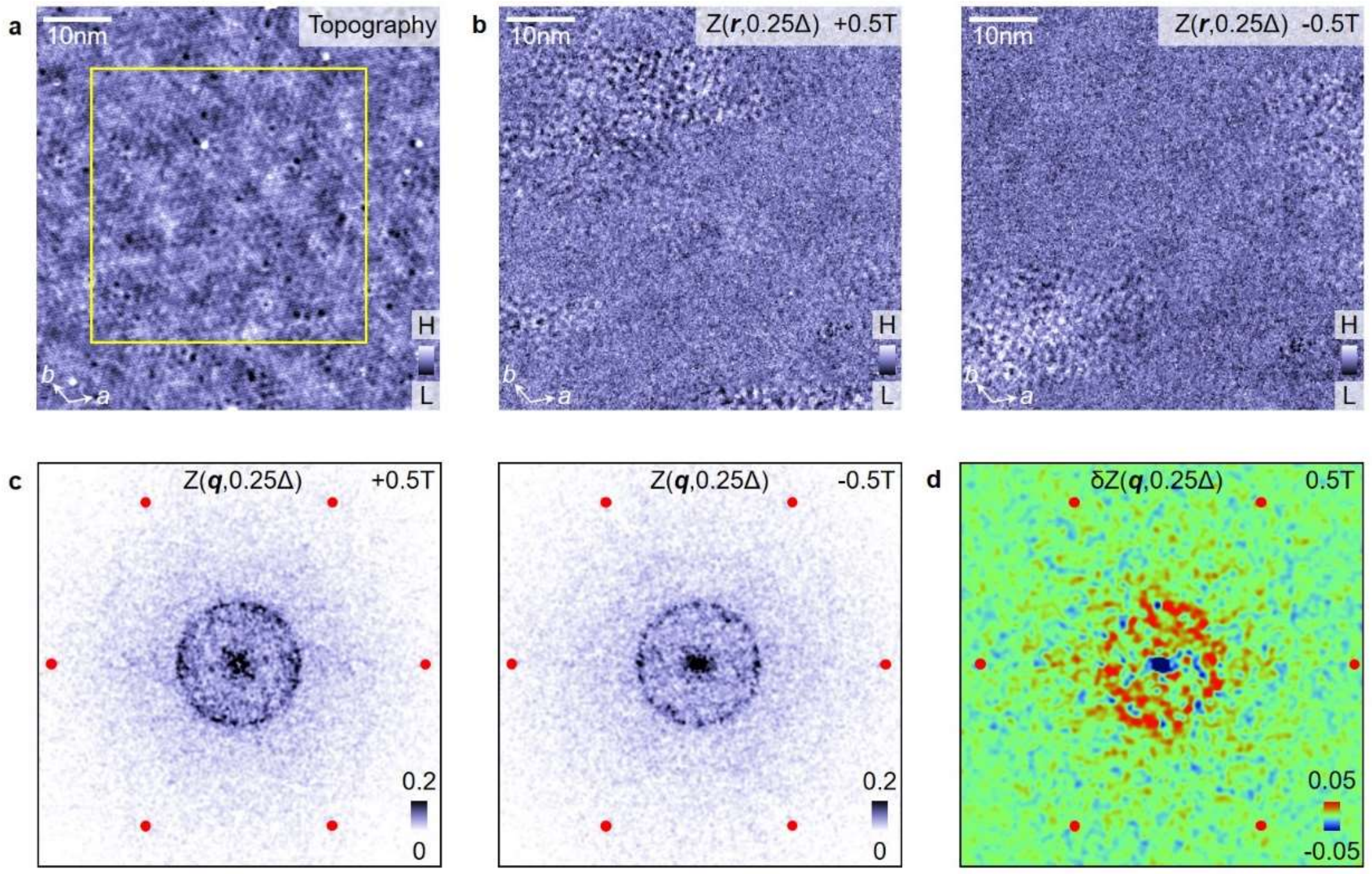

**Figure S16 Time-reversal asymmetrical quasi-particle interference for the gap map region. a,** Topographic image of the related Z maps. The yellow box shows the area of the gap map. **b,**

Corresponding Z(***r***, 0.25Δ) map under magnetic field of ±0.5T applied along the *c*-axis, respectively. **c,** Corresponding Fourier transform of the Z maps in b. **d,** δZ map showing time reversal asymmetrical quasi-particle interference signal at $Q_\alpha$ [δZ=Z(0.5T)-Z(-0.5T)].

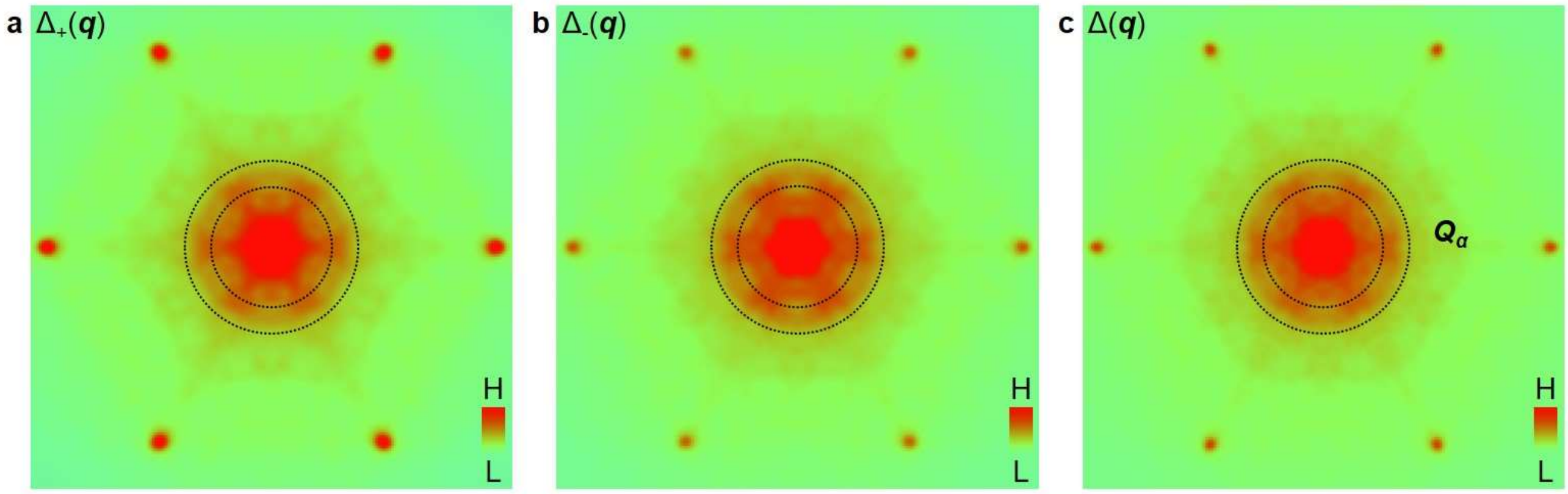

**Figure S17 Robustness of the Fourier transform of the three pair gap maps for the same atomic position.**

## 11. First-principles based orbital analysis

Since the TRSB electronic signal is mainly observed for the $Q_\alpha$, it is meaningful to know the orbital characteristics of the α band from the first principles. First-principles calculations were conducted using the Vienna ab initio Simulation Package[58,59] (VASP) within the Perdew−Burke−Ernzerhof exchange−correlation functional[60]. In our calculations, a kinetic energy cutoff of 520 eV and energy threshold of $10^{-6}$ eV were utilized. To account for van der Waals interaction, the zero-damping DFT-D3 metho[61] was included in our calculations. To simulate the Ta substituted system, the lattice constants $a = b = 5.503$ Å, $c = 9.1414$ Å were used. Band structure and energy contours at Fermi level calculations were conducted using a four quintuple-atomic-layer slab of $CsV_3Sb_5$ with a vacuum layer of 20 Å. Figure S18**a** shows the charge difference for the $V_3Sb_5$ trilayer, which is done by calculating the total charge of the slab and the individual charge of each atom and then subtracting them. The charge difference shows substantial *p-d* hybridizations. Figures S18**b-e** show the orbital projected band structures, identifying the α band is composed mainly of *Sb*-$p_z$ orbital and V $d_{xz/yz}$ orbitals. We further calculate the spectral weight to find the α band is composed of 50% V orbitals, 35% Sb orbitals from the Sb atom within the kagome layer, and 15% Sb orbitals from the Sb atom in the Sb honeycomb layers. It is interesting that the Sb *p* orbitals featuring large spin-orbit coupling contribute substantially to α band. The Sb p orbitals have also recently been discussed to play an important role in the correlation effect[36-40], and their role on TRSB deserves attention.

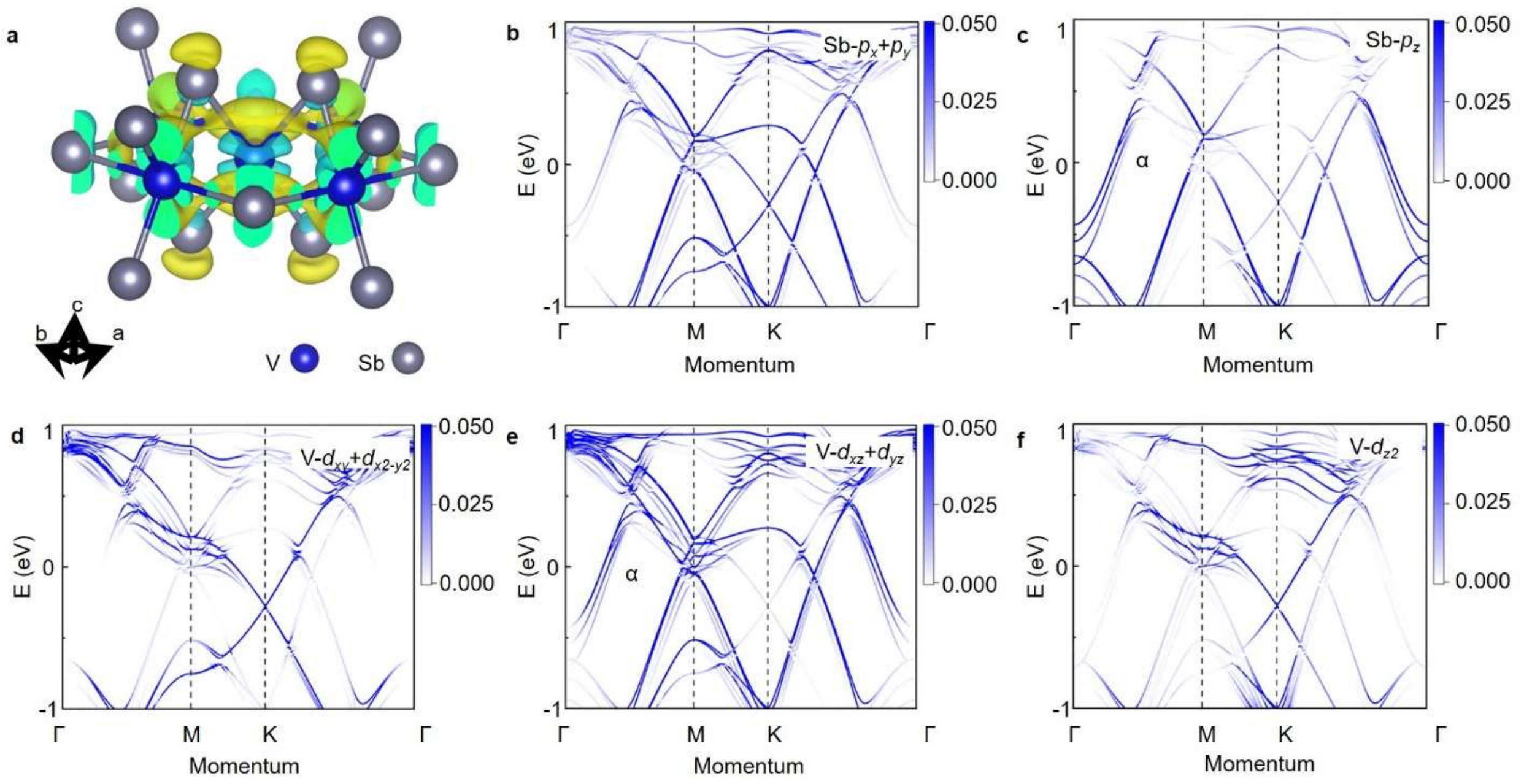

**Figure S18 First-principles based orbital analysis. a,** Charge density difference of $CsV_3Sb_5$. Electron-density increase (decrease) after bonding is shown in yellow (cyan). **b-f,** Orbital projected band structures of the first $Sb_3V_5$ layer for $CsV_3Sb_5$ slab.

## 12. Transverse-field muon spin resonance

In Fig. S19**a**, the transverse-field (TF) μSR-time spectra, measured in a field of = 10 mT, are shown above (T = 5.2 K) and below (T = 0.02 K) the superconducting transition temperature $T_C$. Above $T_C$ the oscillations show a small relaxation due to the random local fields from the nuclear magnetic moments. Below $T_C$ the relaxation rate strongly increases with decreasing temperature due to the presence of a nonuniform local magnetic field distribution as a result of the formation of a flux-line lattice in the Shubnikov phase. Figure S19**b** shows the Fourier transforms of the μSR time spectra shown in Fig. S19**a**. At T = 0.02 K the narrow signal around 10 mT (see Fig. 1**b**) originates from the background, while the broad signal with a first moment slightly smaller than 10mT arises from the superconducting sample. Below $T_C$ a large diamagnetic shift of the internal field $B_{dia}$ between the normal state signal $\mu_0 H_{NS}$ and the superconducting state signal $\mu_0 H_{SC}$ experienced by the muons is observed as shown in Fig. 1**f**. Note that $\mu_0 H_{NS}$ is temperature independent. This diamagnetic shift indicates the bulk character of superconductivity. A fundamental property of the superconducting state that can be directly measured with μSR is the superfluid density. This is accomplished by extracting the second moment of the inhomogeneous field distribution (due to the formation of flux-line lattice) from the muon spin depolarization rate $\sigma_{SC}$, which is related to the superconducting magnetic penetration depth λ as $\langle \Delta B \rangle^2 \propto \sigma_{SC}^2 \propto \lambda^{-4}$ . In order to investigate the symmetry of the superconducting gap, we have therefore derived the temperature-dependent London magnetic penetration depth λ(T), which is related to the relaxation rate by:

$$\sigma_{SC}(T)\,\gamma_\mu = 0.06091\,\Phi_0/\,\lambda^2(T) \qquad (1)$$

Here, $\gamma_\mu$ is the gyromagnetic ratio of the muon, and $\Phi_0$ is the magnetic-flux quantum. The temperature dependence of $\lambda^{-2}$ is shown in Fig. 1**f**. It is clear that $\lambda^{-2}$ (T) reaches its zero-temperature value exponentially, indicating a fully gapped superconductivity.

The raw data of the zero-field μSR data is shown in Fig. S20. We have reported a long stable muon stopping site within $AV_3Sb_5$, located at a distance (3.5 Å) from the vanadium lattice and symmetrically surrounded by a hexagon of vanadium atoms[63]. Muon spin spectroscopy measurements rely on the proper shielding of the sample region from unwanted magnetic fields originating, e.g., from the earth's magnetic field. The zero-field compensation system, which we employed, utilizes the instrument compensation coils with the corresponding currents being dynamically adjusted to account for the time-dependent disturbing fields. To measure these fields, a 3-axis fluxgate probe (Mag-03MC from Bartington Instruments) is permanently mounted near to the sample position. The three analog outputs of the field probe are connected to a data acquisition and control unit. This unit communicates with the μSR slow control system via the MIDAS slow control bus (MSCB). A MIDAS frontend program logs the measured field values and calculates the desired compensation currents that are eventually provided by the control unit. As the position of the 3-axis fluxgate probe is not exactly at the sample position, a careful correction has to be applied to the calculation of the compensation currents to take into account the specific orientation of the probe, the difference between the values of the weak magnetic remanence determined at the probe and sample positions, and the difference between the values of magnetic field components created by the compensation coils pairs at both positions.

This system achieves active zero-field compensation with short- and long-term stability of 1 μT (0.01 G). So the field at the sample is less than 0.01 G. The observed increase in the exponential zero-field muon spin relaxation rate for $Cs(V,Ta)_3Sb_5$ between 5 K and 1.5 K is estimated to be ≈ 0.012 μs-1, which corresponds to a characteristic field strength $\Gamma 12/\gamma_\mu \approx 0.15$ G (which is much higher than 0.01 G). This value could potentially increase if the temperature were lowered to 0.28 K. Therefore, the observed increase is intrinsic and not due to extrinsic effects. The internal field reported for TRSB superconductors (e.g., $Sr_2RuO_4$, $La_7Ni_3$) is typically on the order of 0.1-0.6 G. The similar enhancement of Γ below $T_C$ is observed for $KV_3Sb_5$, $RbV_3Sb_5$, and $CsV_3Sb_5$ under pressure, as shown in Fig. 1f. The zero-field μSR spectra is best described using the gaussian Kubo-Toyabe depolarization function, which reflects the field distribution at the muon site created by the nuclear moments of the sample, multiplied by an additional exponential exp(-Γt) term:

$$P_{ZF}(t) = \left(\frac{1}{3} + \frac{2}{3}(1 - \Delta^2 t^2)\exp\left[-\frac{\Delta^2 t^2}{2}\right]\exp(-\Gamma t)\right) \quad (2)$$

where $\Delta/\gamma_\mu$ is the width of the local field distribution due to the nuclear moments and Γ accounts for the electronic relaxation. The only parameter which exhibits a change as a function of temperature below $T_C$ is the electronic relaxation rate Γ which is plotted in Fig. 1, while Δ is found to be temperature independent. The notable increase of Γ below the onset of superconductivity suggests the appearance of a spontaneous magnetic field, providing evidence of TRSB on entering the superconducting state. We also note that the nuclear moment is large in this material, and the relaxation rate Δ at 15K is already $0.27\mu s^{-1}$. When we see increase of the electronic relaxation rate on top of the large nuclear rate, the relative difference seen by eye is small.

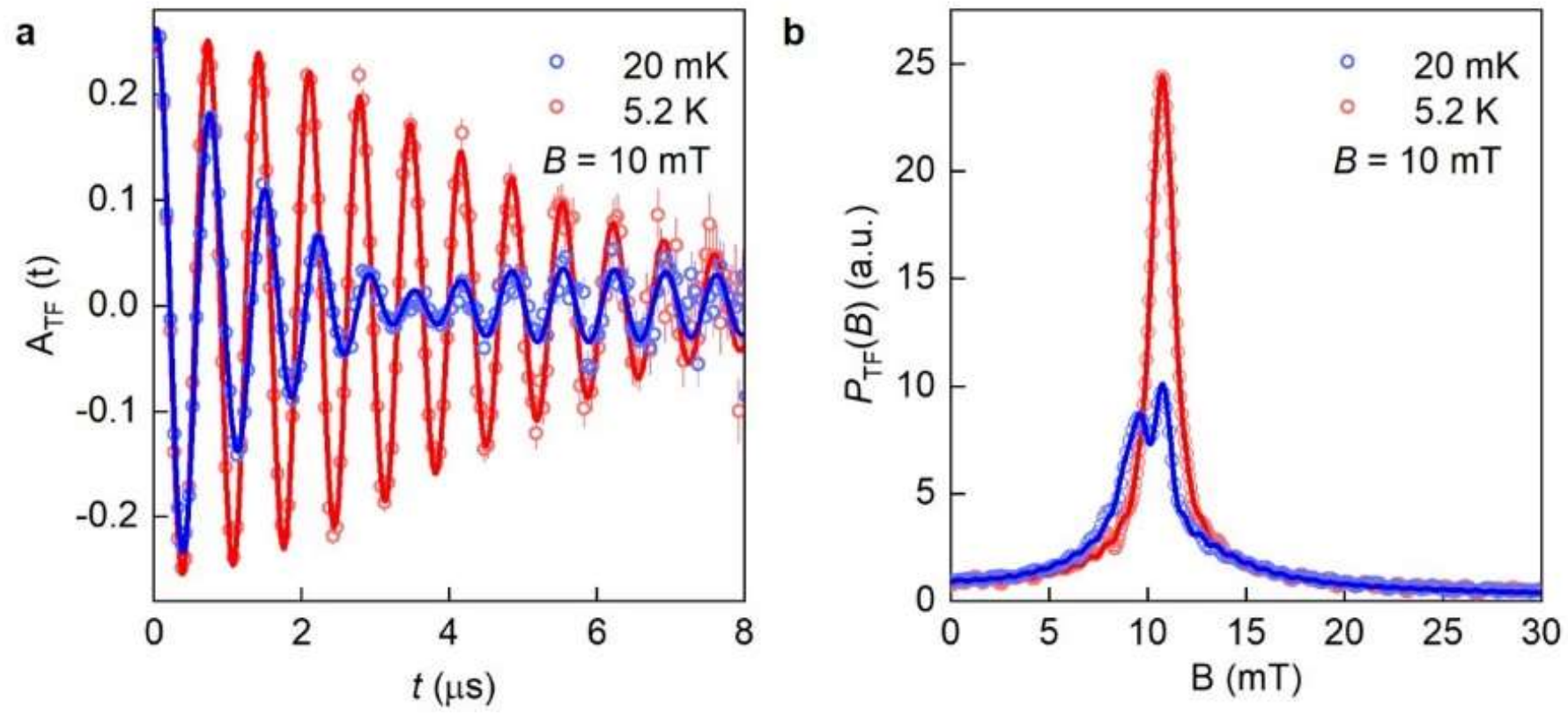

**Figure S19 Transverse field muon spin resonance data.** Transverse-field (TF) μSR time spectra **a,** and the corresponding Fourier transforms **b,** measured in an applied magnetic field of 10 mT. The spectra above (5.2 K) and below (0.02 K) the superconducting transition temperature Tc (after field cooling the sample from above Tc) are depicted. The solid lines represent fits to the data using the following functional form $A_{TF}(t)= \sum_{i=1}^{2} A_i \exp\left[-\frac{\sigma_i^2 t^2}{2}\right] \cos(\gamma_\mu B_{int,i} t + \varphi)$. Two component expression was used due to the observed weak asymmetric field distribution. $A_i$, $B_{int,i}$, and $\sigma_i$ are the symmetry, the mean field and the relaxation rate of the $i$th component, and $\varphi$ is the initial phase of the muon spin ensemble. $\gamma_\mu$=135.5 MHz $T^{-1}$ is the muon gyromagnetic ration. In order to extract the second moment of the field distribution from the two-component fitting, we used a similar procedure as described in Ref. 62.

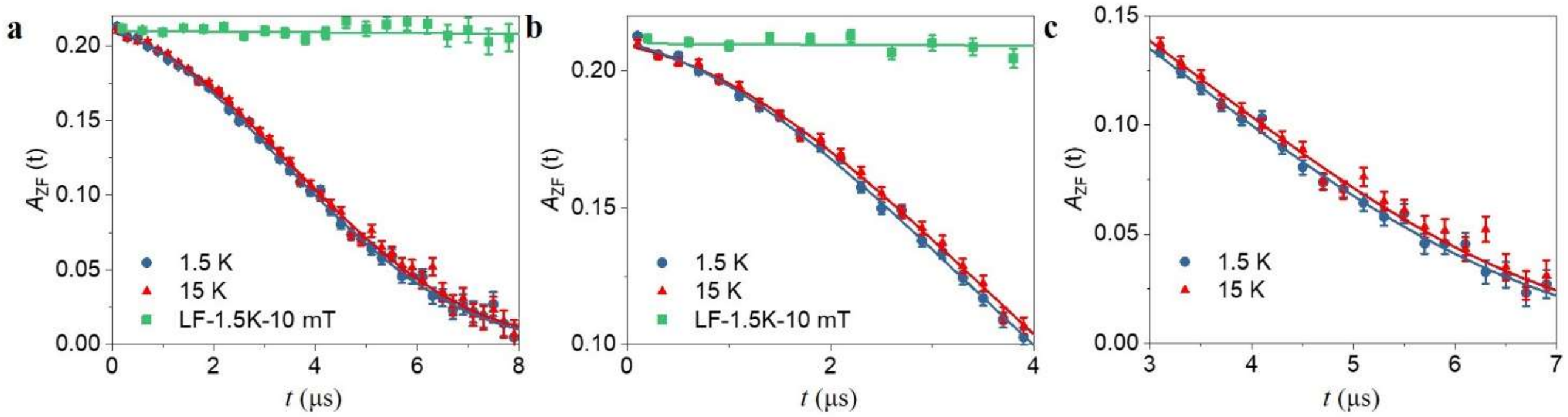

**Figure S20 Zero-field μSR data. a,** The μSR time spectra measured in zero-field above and below $T_C$ and under a small external magnetic field applied in a direction longitudinal to the muon spin polarization, $B_{LF}$ = 10 mT. **b,c,** A zoomed-in view of the low and high time segments of the data is provided to illustrate the differences in the spectra between the normal and superconducting states.

## 13. Polarized neutron scattering experiments: absence of long-range magnetic order and moment constraints for TRSB superconductivity

Our polarized neutron experiments were carried out on the CEA-CRG triple axis spectrometer IN22 at the Institut Laue-Langevin (ILL), Grenoble, France and HB-1 polarized triple axis spectrometer of the High Flux Isotope Reactor (HFIR), at Oak Ridge National Laboratory (ONRL), USA. The sample was aligned with the [110] and [001] wave vectors within the horizontal scattering plane giving access to the (*H*, *H*, *L*) Bragg reflections. Figure S21**b** shows a rocking scan of the sample assembly, indicating an effective sample mosaicity of about 1.5 degrees with subsequent weaker peaks. The hexagonal lattice parameters of the sample assembly were measured to be $a$ = 5.503 Å and $c$ = 9.1414 Å.

The longitudinal Polarization analysis measurements on IN22 were carried using standard XYZ-Helmholtz-like coils to control the polarization $P_i$ of the incident neutron beam and to analyze the longitudinal components of $P_f$, the polarization of the scattered beam. In this work, the X axis of the reference frame is defined to be parallel to the momentum transfer $Q = k_i - k_f$, the Z axis is perpendicular to the scattering plane, and Y axis forms a right-handed system (Fig. S22**c**). We measured the spin-flip (SF) and non-spin-flip (NSF) intensities. An isotropic flipping ratio along X, Y, and Z, $R = I_{SF}/I_{NSF} = 16.2 \pm 0.3$, was estimated from the peak intensity of the Bragg peak (1,1,0) in the normal state (Fig. S21**d**). The depolarization technique was used to determine the superconducting temperature of the sample assembly[64]. The spectrometer was set for the (1,1,0) nuclear Bragg reflection. The sample was neutron guide field-cooled through the superconducting transition with an applied field of a few Gauss along the Y direction and then rotated by 90° towards the X direction. The flipping ratio is noticeably reduced by the trapped flux in the sample due to superconductivity that depolarizes the beam and is restored to its nominal value by warming up across the superconducting temperature. A superconducting transition at 5 K is found for the entire bulk of the sample in agreement with the resistivity measurement of individual single crystals.

To evidence any sign of ferromagnetic order in the sample, we measured the SF and NSF intensities at the Bragg peak (1,1,1) at 6 K at IN22 and (1,1,0) at 1.5 K and 7 K at HB-1. No magnetic peak is observed across the Bragg peak in the Spin-flip channel in both the normal state and superconducting state. Table S1 and Table S2 show the SF intensity at the Bragg position of (1,1,1) and (1,1,0), respectively. $I_c = I^X_{SF} - I^Y_{SF}$ corresponds mainly to the magnetic intensity from the magnetic moment along the c-axis and $I_a = I^X_{SF} - I^Z_{SF}$ corresponds to the magnetic intensity for the magnetic moment in-plane. $I_{mag} = 2I^X_{SF} - I^Y_{SF} - I^Z_{SF}$ represents the total measured magnetic intensity[64]. Within error bars, no sign of ferromagnetic order is observed in any of these intensity differences both above and below the superconducting transition temperature.

To estimate the upper limit of the possible magnetic intensity, we calibrated the magnetic intensities in absolute units by scaling the intensities to the strong nuclear Bragg peak (1,1,0)[65]. The nuclear structure factor of (1,1,0) is $|F(1,1,0)|^2$ = 10.65 barn using the VESTA software. The measured Bragg peak intensity is 2.4e5 for the same monitor counts. Corrections from resolution effects are negligible here as both Bragg positions are close to each other in reciprocal space. Using the above numbers, the upper limit of the ferromagnetic intensity of $I_c$, $I_a$, and $I_{mag}$ at 6 K are deduced to be 0.11, 0.11, and 0.18 mbarn, respectively.

Furthermore, from magnetic intensities, one can estimate the upper limit of the magnetic moment using $I_{mag} = 290|F(Q)|^2M^2$, where $|F(Q)|^2$ is the magnetic form factor and M the possible magnetic moment in Bohr magneton ($\mu_B$) units per formula unit[64]. We take the magnetic form factor of $V^{4+}$ atom at Q = (1,1,1) as an estimate, which is $|F(Q)|^2$ =0.49. This leads to an upper limit of the magnetic moment of

$M < 0.035\ \mu_B$ for the total magnetic moment and $M_c < 0.028\ \mu_B$ for a moment along the c-axis per formula unit at 6 K. Making the assumption that the magnetic moments are only carried by the vanadium atoms, it gives a moment of less than 0.01 $\mu_B$ per vanadium atoms.

In addition to measurements at the Bragg wave vectors, we also investigated the charge order wavevector (0.5, 0.5, 0) both above and below the superconducting transition temperature at IN22. Table S3 shows the SF intensities at (0.5, 0.5, 0). When the sample is in the normal state, no sign of magnetism is observed at (0.5 0.5, 0) within error bars. Following the recipe for the calibration of the Bragg (111) peak and considering additional correction of the Cooper-Nathans-type resolution function, one can estimate the magnetic contribution at (0.5, 0.5, 0) from magnetic moment along the crystalline c axis. When the sample is in the superconducting state, $I_{mag} = 0.08 \pm 0.07$ mbarns. To estimate the magnitude of magnetic moment, a specific magnetic model is needed. A variety of chiral flux phase models have been proposed to explain the complex charge-ordered states in $CsV_3Sb_5$ related metals[13,53-57]. In the 2×2 charge order unit cell of loop current model in Ref.54, one can view it as one moment up $M_1$ (along +c axis) at position (0.5, 1.5) and three moments down $M_2$ (along -c axis) at positions (0.5, 0.5), (1.5, 0.5), and (1.5, 1.5). Since there is no net moment per unit cell, $M_1+3M_2=0$. Thus, the corresponding magnetic structure factor at (0.5, 0.5, 0) is $|M(Q)|^2 = 16/9\ M_1^2$. Then for one unit cell, one has $I_{mag} = 290\times1/4\times|F(Q)|^2\times16/9\times M_1^2$. Applying the relation above with an estimation of form factor $|F(Q)|^2 = 0.75$, the magnetic moment $M_1$ is estimated to be about $0.03 \pm 0.02$ μB. These polarized neutron experiments and related estimations set up further constraints for the future modelling of the TRSB superconductivity.

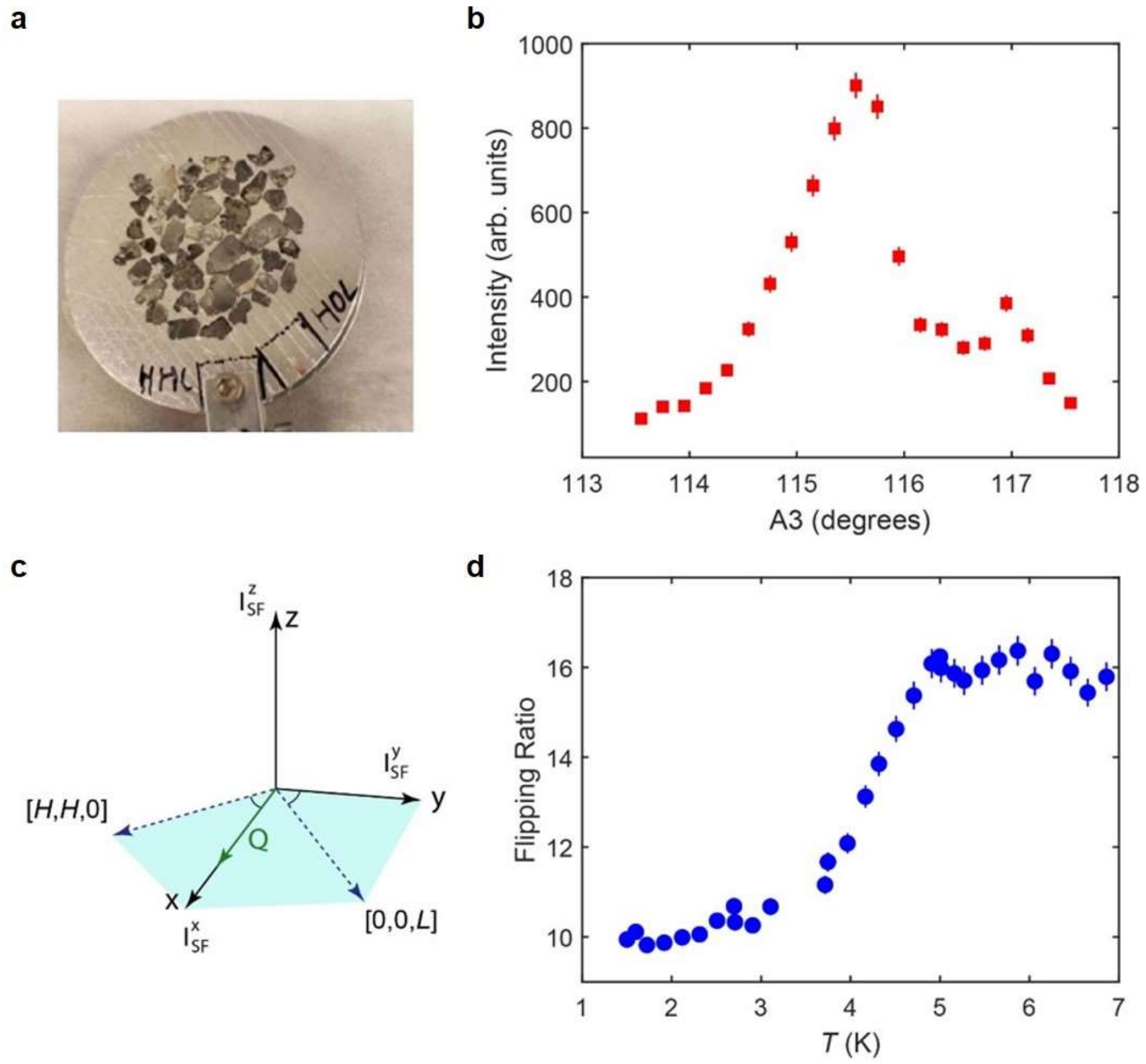

**Figure S21 Polarized neutron scattering experiment. a,** Picture of the sample assembly of $Cs(V_{0.86}Ta_{0.14})_3Sb_5$ single crystals. **b,** Neutron rocking scan on the Bragg peak (1,1,0) of the sample

assembly. **c,** Scattering geometry of polarized neutron scattering experiment in the [*H*,*H*,*L*] plane. **d,** Flipping ratio measured by neutron depolarization that shows a superconducting transition at 5K of the sample assembly.

| $I^X_{SF}$ | $I^Y_{SF}$ | $I^Z_{SF}$ | $I_{mag} = 2I^X_{SF}-I^Y_{SF}-I^Z_{SF}$ | $I_c = I^X_{SF}-I^Y_{SF}$ | $I_a = I^X_{SF}-I^Z_{SF}$ |
|---|---|---|---|---|---|
| 90.1 ±1.3 | 90.6 ±1.9 | 88.9 ±1.9 | 0.7 ±3.8 | -0.5 ±2.3 | 1.2 ±2.3 |

**Table S1**, Spin-flip intensities at the Bragg peak (1,1,1) at 6 K for a monitor M = 1e6 (corresponding to a counting time of 237 sec).

| *T* (K) | $I^X_{SF}$ | $I^Y_{SF}$ | $I^Z_{SF}$ | $I_{mag} = 2I^X_{SF}-I^Y_{SF}-I^Z_{SF}$ | $I_c = I^X_{SF}-I^Y_{SF}$ | $I_a = I^X_{SF}-I^Z_{SF}$ |
|---|---|---|---|---|---|---|
| 7 | 3008.0 ±17.3 | 3060.3 ±17.5 | 2975.0 ±17.3 | -19.2 ± 42.5 | -52.2 ± 24.6 | 33.0 ± 24.5 |
| 1.5 | 2993.9 ±10.0 | 3037.2 ±10.1 | 2999.9 ±10.0 | -49.2 ± 24.5 | -43.3 ± 14.2 | -5.9 ± 14.1 |

**Table S2**, Spin-flip intensities at the Bragg peak (1,1,0) at 7 K and 1.5 K for a counting time of 1 minute.

| *T* (K) | $I^X_{SF}$ | $I^Y_{SF}$ | $I^Z_{SF}$ | $I_{mag} = 2I^X_{SF}-I^Y_{SF}-I^Z_{SF}$ | $I_c = I^X_{SF}-I^Y_{SF}$ | $I_a = I^X_{SF}-I^Z_{SF}$ |
|---|---|---|---|---|---|---|
| 6 | 99.1 ± 1.6 | 98.6 ± 2.3 | 99.2 ±2.3 | 0.3 ±4.6 | 0.4 ±2.8 | -0.1 ±2.8 |
| 1.5 | 100.1 ± 1.7 | 96.9 ± 2.3 | | | 3.2 ±2.8 | |

**Table S3**, Spin-flip intensities at the CDW position (0.5, 0.5, 0) for a monitor M = 1e6 (corresponding to a counting time of 237 sec).

**14. Extended discussions on the QPI experiment**

We have found that there is a significant difference between QPI with magnetic fields pointing in opposite directions which suggests that the full gap superconductivity in this kagome metal may host a complex superconducting order parameter $\Delta_1+i\Delta_2$ (such as s+is, p+ip, and d+id) that breaks time-reversal symmetry. For an applied field, the superconductor enters into the vortex state where individual vortices act as additional disorders and contribute to the QPI intensity. In a simple picture, the vortex is a combination of scattering from a reduced order parameter in the vortex core and magnetic scattering from the flux line since there is a finite Zeeman field inside the vortex that can be considered as an extended magnetic impurity. The first contribution is independent of the direction of the external magnetic field and the direction of the time-reversal symmetry breaking of the order parameter and will therefore drop out when the difference δZ(B,T) is calculated. Scattering on magnetic impurities is however different for the case of the magnetic field +B and -B in a superconductor that breaks time-reversal symmetry ($\Delta_1+i\Delta_2$). This results in the signal that is picked up when examining the experimental data δZ(B,T). Indeed, above $T_C$, there are no vortices and therefore no additional QPI signal. This is in line with the disappearance of δZ(B,T); the same is true for energies beyond the superconducting gap where the electronic structure essentially is identical to the band structure in the normal state and scattering on effective magnetic impurities (from the vortices) is identical for fields +B and -B as observed in the experimental data. In summary, the time-reversal asymmetrical interference of Bogoliubov quasi-particles points towards a complex superconducting order parameter whose complete determination requires further model-dependent theoretical analysis in reference to our experiments.